\documentclass[11pt]{article}
\usepackage{fullpage}
\usepackage{amsmath}
\usepackage{amssymb}
\usepackage{amsthm}
\usepackage{amsfonts}
\usepackage{bm}
\usepackage{enumitem}
\usepackage[final]{graphicx}
\usepackage{caption}
\usepackage{float}
\usepackage{epstopdf}
\usepackage{subcaption}
\usepackage{multirow}
\usepackage{multicol}
\usepackage{booktabs}
\usepackage{algorithm}
\usepackage{algpseudocode}
\usepackage{url}
\usepackage{cite}
\usepackage{textcomp}
\usepackage{subcaption}
\usepackage[font=small,labelfont=bf]{caption}
\usepackage{multirow}
\usepackage{rotating}
\usepackage{microtype}
\usepackage{authblk}
\usepackage{url}
\usepackage{titlesec}

\author[*]{Siddharth Patel}
\author[*$\S$]{Mohammad Rasouli}
\author[$\dag$]{Junjie Qin}
\author[*$\ddag$]{Ram Rajagopal}
\affil[*]{Department of Civil and Environmental Engineering, Stanford University}
\affil[$\S$]{Department of Management Science and Engineering, Stanford University}
\affil[$\dag$]{Department of Electrical Engineering and Computer Science, University of California, Berkeley}
\affil[$\ddag$]{Department of Electrical Engineering (by Courtesy), Stanford University}

\date{}

\title{The Value of Distributed Energy Resources for Heterogeneous Residential Consumers}

\begin{document}

\maketitle

\begin{abstract}
The presence of behind-the-meter rooftop photovoltaics and storage in the residential sector is poised to increase significantly.
Here we quantify in detail the value of these technologies to consumers and service providers.
We characterize the heterogeneity in household electricity cost savings under time-varying prices due to consumption behavior differences. 
Different pricing policies significantly alter how households fare with respect to one another.
Furthermore, household savings in absolute terms are not strongly correlated with savings normalized by PV and storage system size.
We characterize the financial value of improved forecasting capabilities for a household, finding that it is a relatively small fraction of a household's cost savings.
Coordination services that combine the resources available at all households can reduce costs by an additional 10\% to 15\% of the original total cost.
Surprisingly, coordination service providers will not encourage adoption beyond 35-55\% within a group.
We present a simple model that explains the value of coordination and its relationship to the pricing of distribution services.
\end{abstract}

Distributed energy resources (DERs) are an essential part of modernizing and de-carbonizing the grid \cite{casa2013,akorede2010,jain2017} and pose challenges for the design, management, and operation of the electricity system \cite{kok2016,ferreira2013,xu2016}.
Dramatic changes are expected as consumers adopt behind-the-meter DERs and become prosumers capable of responding to prices and other signals from grid operators \cite{parag2016,bayram2017}.
Technology vendors (e.g. Tesla) and DER resource aggregators (e.g. OhmConnect) will play a significant role in the emerging ecosystem by making DERs accessible to smaller consumers and ensuring that they operate those technologies in a manner aligned with their self-interest and compatible with the needs of the grid as a system.
The proliferation of DERs will create opportunities for new business models as well.
The coming impact of the adoption of energy storage and rooftop photovoltaic (PV) systems by residential consumers is not well understood\cite{agnew2015}.

The impact of DERs depends on the adoption rate of technology, the operation strategy, and the financial and policy arrangements for system participants.
Consumer behavior and the resulting consumption pattern heterogeneity are critical drivers of DER value and govern the interactions between these dimensions.
Residential electricity consumption exhibits significantly more diversity than previously believed \cite{kwac2014}.
Yet this heterogeneity is seldom accounted for in existing studies, which use data from a small number of residential or commercial consumers to evaluate the consumer-side economics of PV \cite{darghouth2011,borenstein2007,bright2009,mills2008,vangeet2008}, storage \cite{wu2016,peterson2010}, or both combined \cite{nottrott2012,su2001,stadler2008,stadler2009,neubauer2015,hoff2007,riffonneau2011}.

In this paper we propose a simple and scalable methodology to estimate DER value and impact that incorporates consumption heterogeneity.
We apply our methodology to provide a first of its kind assessment of the value of these technologies to residential consumers, technology vendors, and aggregators under various business models and policy arrangements.
We focus on PV and storage because they are commercially available and gaining traction as their costs decrease \cite{eia2015,eia2016,deutsche2016,nykvist2015,bronski2015}.
We also note that public entities in California have adopted goals and standards requiring that all new homes have PV systems, whether individually or as part of community solar, starting in 2020 \cite{CEC2015,CECRelease2018}.
Furthermore, federal tax policy in the United States encourages storage adoption when co-located with and charged by PV \cite{comello2018,hinds2014}.
Our study accounts for adoption rates and consumption heterogeneity in unique ways to identify tipping points for impact.
We utilize a large dataset of hourly electrical energy consumption recordings for residential consumers in Northern California to capture heterogeneity.
Based on this data we build models to assess the value of storage and PV to consumers in the form of bill savings.

We estimate the value that can be delivered by entities that provide improved information to and enable resource sharing among residential consumers.
Households rely on information about future consumption and generation when deciding how to operate their DERs.
We model how constraints on this information impact the value of DERs to households.
This analysis gives an estimate of the value of services that improve the accuracy of information available to the households.
Communication technology enables aggregators to coordinate and share DERs among a group of households.
We model and characterize the value that these coordination services can generate, which depends on the pattern of technology adoption by households, the sharing mechanism, and the pricing policy.

\section{Assessing value from data}

Storage devices enable households to shift their energy usage in time, and rooftop PV generates electricity that households can consume directly or sell back to the grid.
Both technologies allow households to reduce their electricity costs.
We estimate how much households and groups of households could save on their bills if they adopted a rooftop PV system sized to make the household net-zero in terms of electrical energy, along with a storage device with capacity scaled to the PV system size at 1 kWh capacity for every 1 kW of PV system size.
For example, if a household's net-zero PV system is sized at 5 kW, then its storage device will have capacity 5 kWh.
We choose net-zero sizing because it is related to explicit policy goals being advanced in the United States and Europe \cite{CEC2015,CECRelease2018,sartori2012,annunziata2013}.
We assume the households operate these devices to minimize their electricity costs.
We take a household's hourly smart meter data as its inflexible end-use consumption.

We provide an \textit{a-priori} snapshot of potential bill savings.
Thus, electricity prices are exogenous in our model.
Similarly, we assume that households do not significantly alter their electricity consumption behavior when they adopt DERs or when they face differing rates.
Refer to the Methods section for detailed explanations of the data sources and analyses.

\begin{figure}
  \begin{minipage}[]{.4\linewidth}
    \centering
    \includegraphics[width=.9\linewidth]{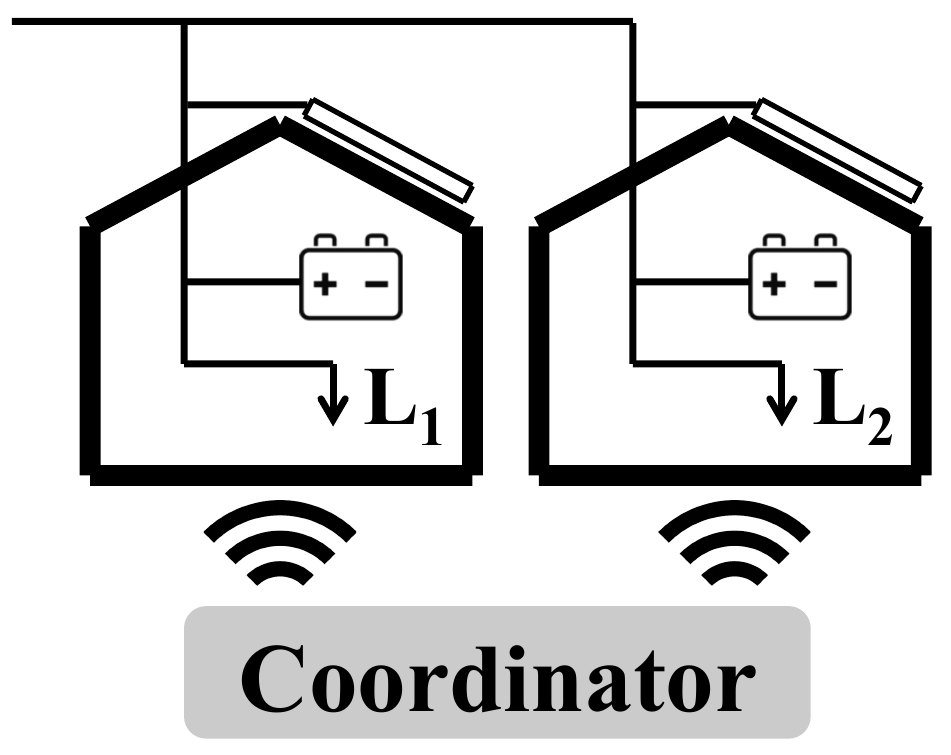}
    \scriptsize
    \centering \\
    \hspace{8pt}
\begin{tabular}{c|l|l}
Policy  & Purchase Price & Sale Price            \\ \hline
1 & Retail TOU     & Wholesale             \\ \hline
2 & Retail dynamic      & Discounted dynamic             \\ \hline
3 & Retail TOU     & Discounted TOU \\ \hline
4 & Flipped TOU & No selling
\end{tabular}
  \end{minipage}%
  \begin{minipage}[]{0.60\linewidth}
\begin{figure}[H] 
  \begin{subfigure}[b]{0.33\linewidth}
    \centering
    \includegraphics[width=0.9\linewidth]{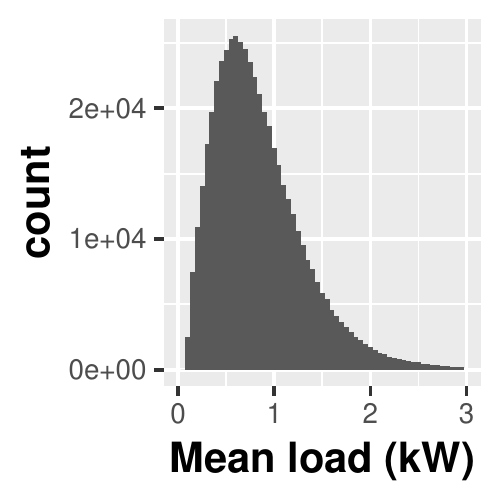} 
    \caption{} 
    \label{fig:Mean_load} 
    \vspace{0pt}
  \end{subfigure}
  \begin{subfigure}[b]{0.33\linewidth}
    \centering
    \includegraphics[width=0.9\linewidth]{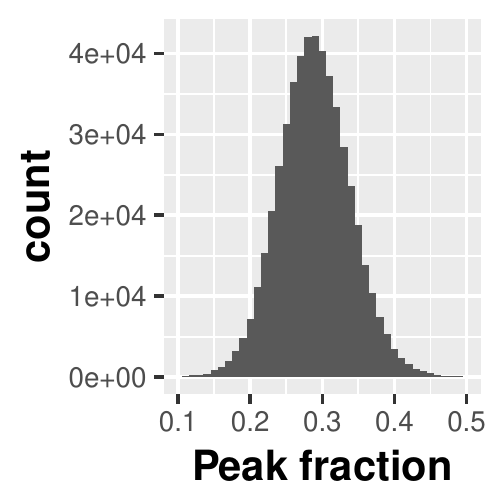} 
    \caption{} 
    \label{fig:Peak_fraction} 
    \vspace{0pt}
  \end{subfigure}
  \begin{subfigure}[b]{0.33\linewidth}
    \centering
    \includegraphics[width=0.9\linewidth]{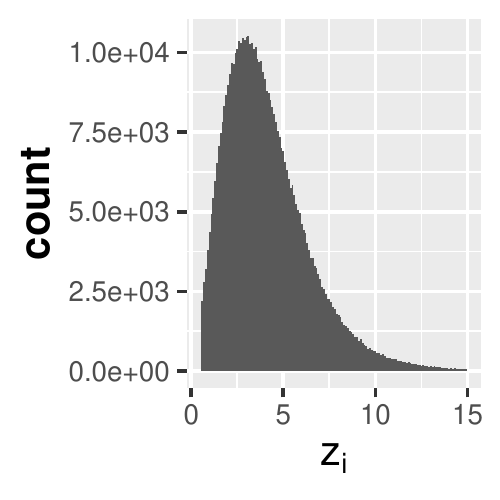} 
    \caption{} 
    \label{fig:z_i} 
    \vspace{0pt}
  \end{subfigure}
  \begin{subfigure}[b]{0.33\linewidth}
    \centering
    \includegraphics[width=0.9\linewidth]{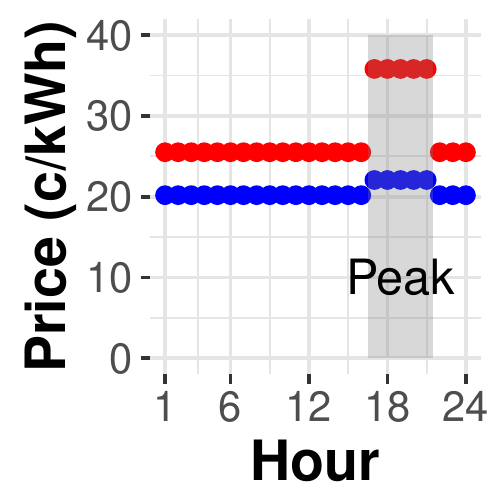} 
    \caption{} 
    \label{fig:TOU} 
  \end{subfigure}
  \begin{subfigure}[b]{0.33\linewidth}
    \centering
    \includegraphics[width=0.9\linewidth]{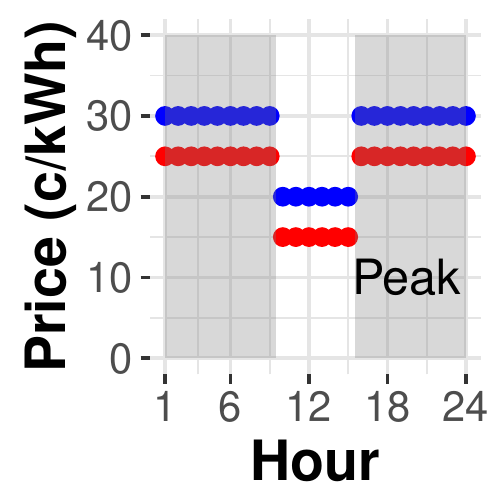} 
    \caption{} 
    \label{fig:flipped_TOU} 
  \end{subfigure}
  \begin{subfigure}[b]{0.33\linewidth}
    \centering
    \includegraphics[width=0.9\linewidth]{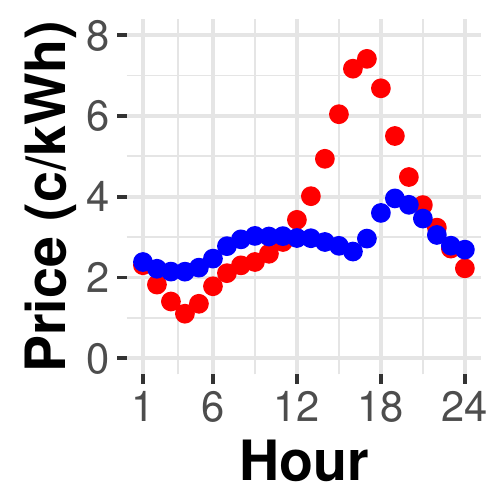} 
    \caption{} 
    \label{fig:LMP} 
  \end{subfigure} 
  \label{fig:Simulation_setup} 
\end{figure}
\end{minipage}%
\caption{There is considerable heterogeneity in the households in our dataset, which together consume 3.9 GWh of electrical energy annually and have a peak load of 1.2 MW. Fig. (a) shows the distribution of mean electrical load for the households in our dataset (range 0.1-20.1 kW), Fig. (b) shows the distribution of the fraction of consumption that takes place during peak hours of the retail TOU rate (range 0.00-0.73), and Fig. (c) gives the distribution of $z_i$, the net-zero PV system size in kW (range 0.5-100.7). In our study, each household has a given inflexible load, $\textbf{L}_i$, a rooftop PV system sized at $z_i$ kW, and a storage device with capacity $z_i$ kWh. The PV and storage are connected to the house AC bus through inverters. The table defines pricing policies used in the simulations. Fig. (d) plots the retail TOU rate for a summer (red) and winter (blue) business day. Fig. (e) does the same for the flipped TOU rate, which serves as a speculative worst-case for disincentivizing rooftop PV. Fig. (f) plots example days for the wholesale rate.  The retail dynamic rate is a scaled version of the wholesale rate, exposing households directly to wholesale market price variations.}
\label{fig:Diagram_and_Scenarios}
\end{figure}

\section{Value of technology for households}

We define absolute annual savings for a household, $S_{a,i}$, as how much its electricity bill decreases when it adopts PV and storage.
Let $z_i$ be the kW rating of the home's net-zero PV system and therefore also the kWh capacity of its storage device.
A household's normalized savings is $S_{n,i}=S_{a,i}/z_i \textrm{ kW-kWh}$, its absolute annual savings divided by its system size.
This gives a \$/kW-kWh/year savings estimate for each household, where the kW-kWh unit denotes that the system consists of both PV and storage.
The normalized savings can be compared directly across households because it accounts for varying net-zero system sizes.
Normalized savings also can be compared directly to annualized per-unit system costs.
We compute savings under four pricing policies, incorporating time-of-use (TOU), wholesale, and dynamic rates, as listed in the table in Figure  \ref{fig:Diagram_and_Scenarios}.
The TOU and discounted TOU rates are the same for all households, whereas the wholesale and dynamic rates vary for households based on location.

\subsection{Heterogeneity in savings}

\begin{figure}
        \centering
        \begin{subfigure}[b]{0.475\textwidth}
            \centering
            \includegraphics[width=\textwidth]{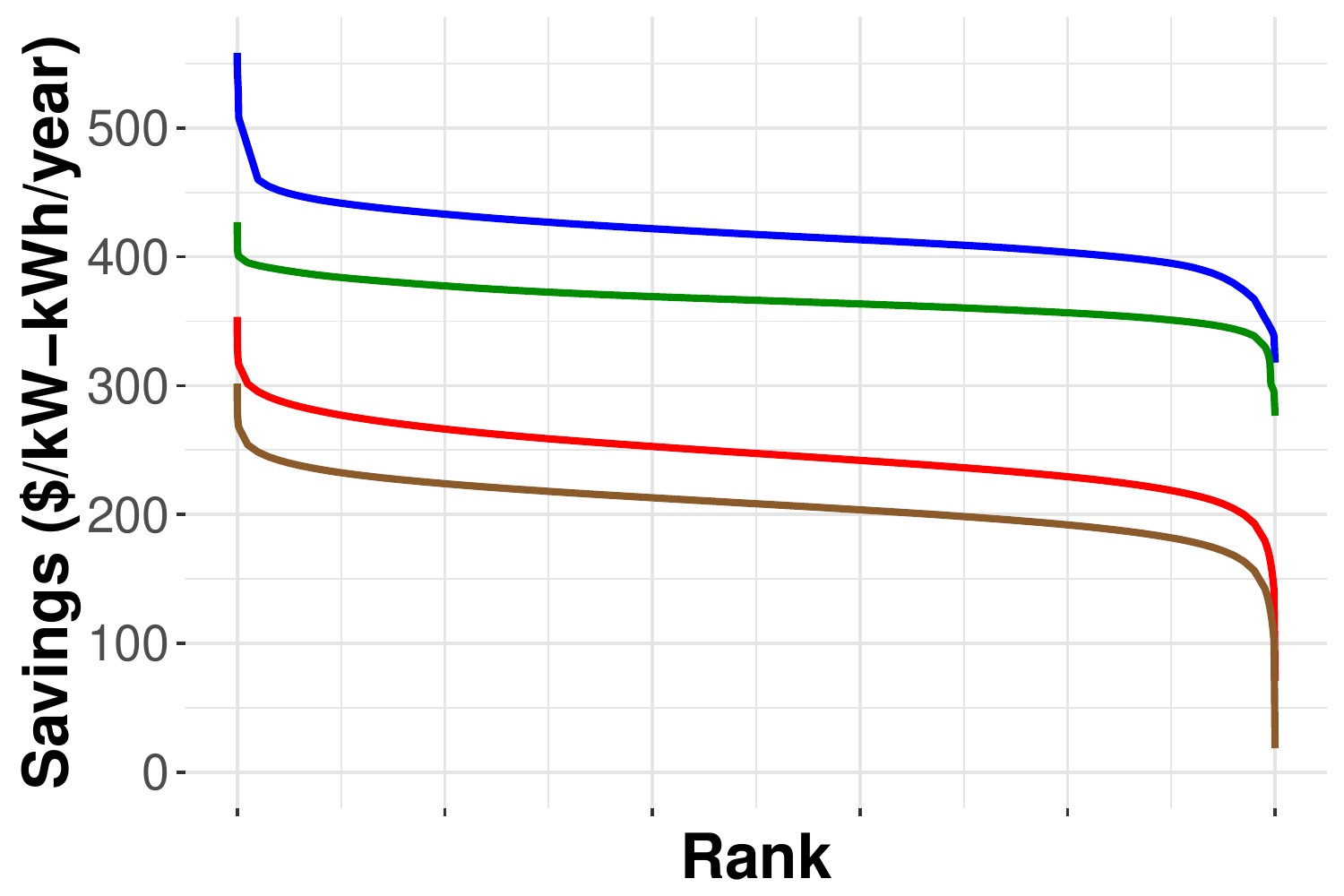}
            \caption{}
            \label{fig:Normalized_Savings}
        \end{subfigure}
        \hfill
        \begin{subfigure}[b]{0.475\textwidth}  
            \centering 
            \includegraphics[width=\textwidth]{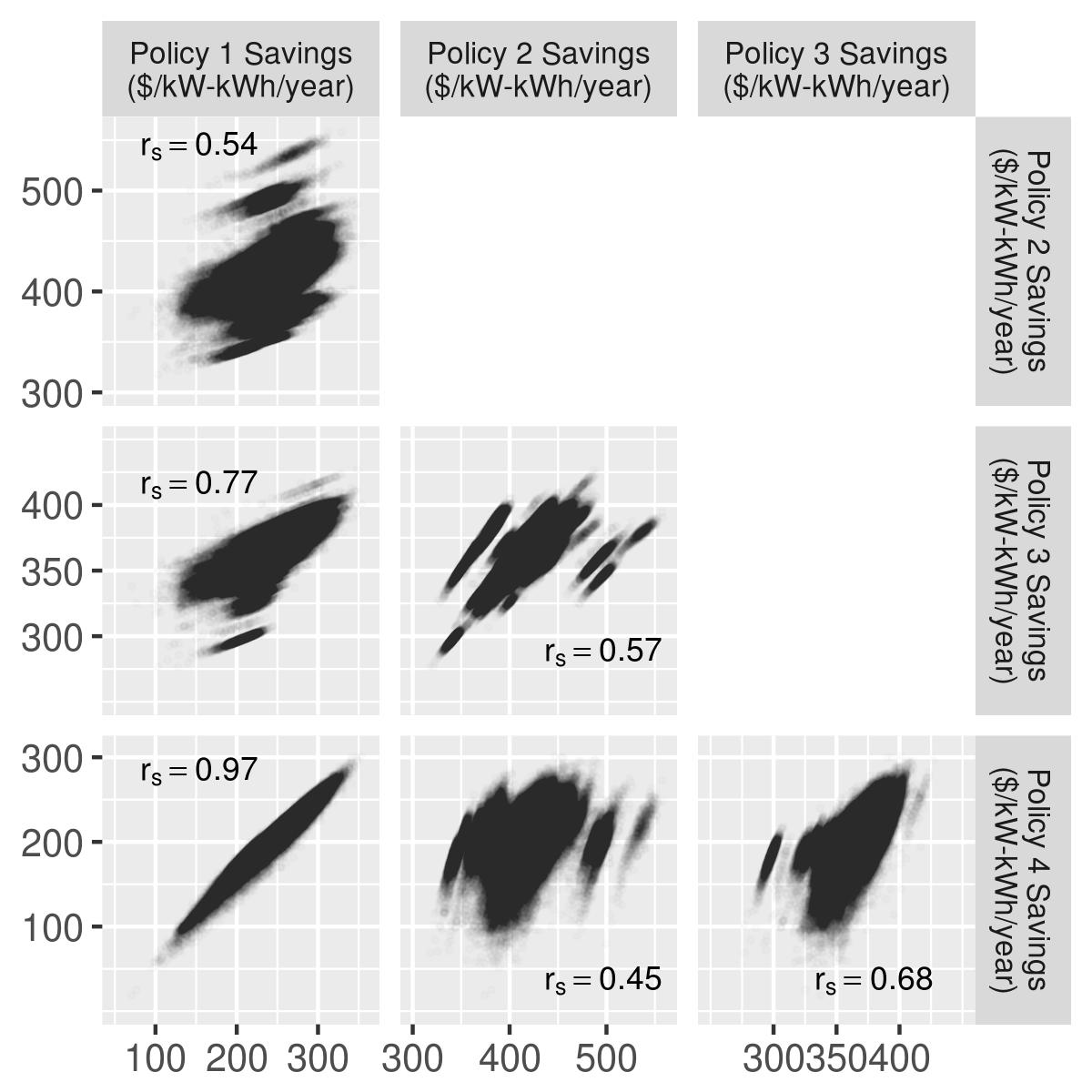}
            \caption{}  
            \label{fig:Pairs_sysnorm}
        \end{subfigure}
         \vskip\baselineskip

\begin{subfigure}[b]{0.2375\textwidth}   
             \centering              \includegraphics[width=\textwidth]{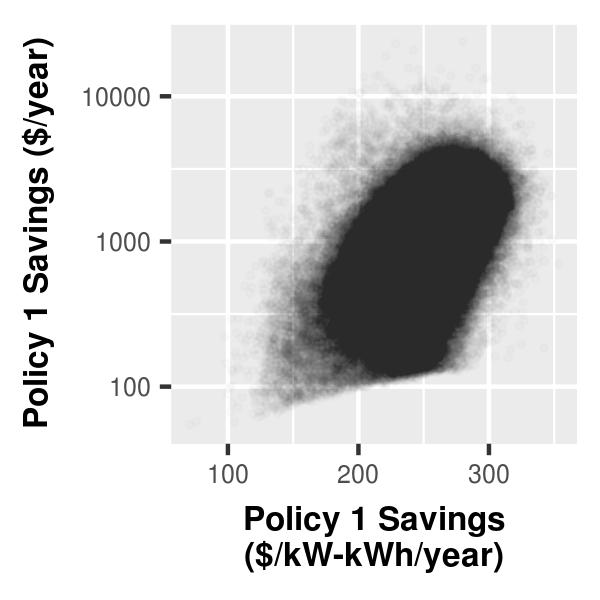}
             \caption{}  
             \label{fig:p1_abs_v_norm}
         \end{subfigure}
\begin{subfigure}[b]{0.2375\textwidth}   
             \centering
\includegraphics[width=\textwidth]{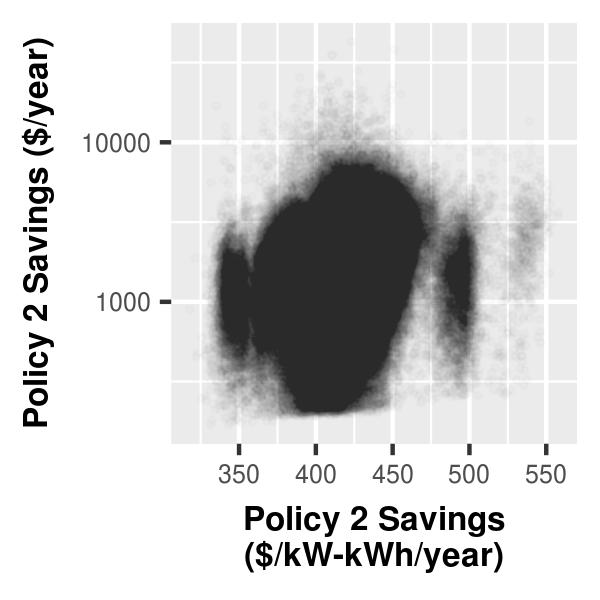}
             \caption{}  
             \label{fig:p2_abs_v_norm}
         \end{subfigure}
\begin{subfigure}[b]{0.2375\textwidth}   
             \centering 
\includegraphics[width=\textwidth]{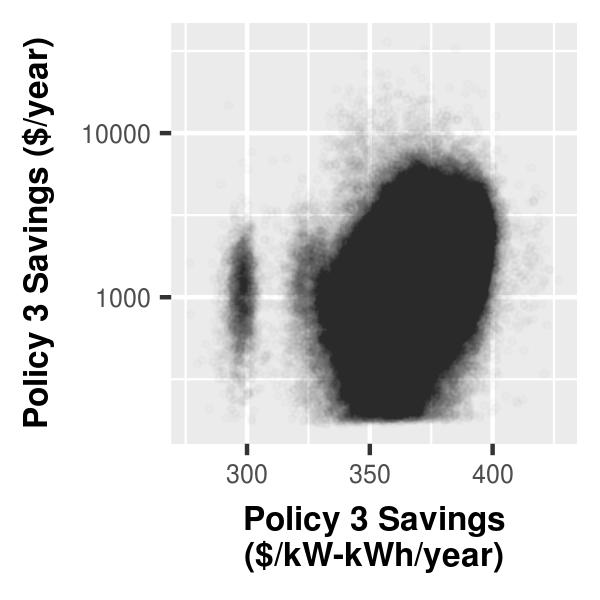}
             \caption{}  
             \label{fig:p3_abs_v_norm}
         \end{subfigure}
\begin{subfigure}[b]{0.2375\textwidth}   
             \centering 
\includegraphics[width=\textwidth]{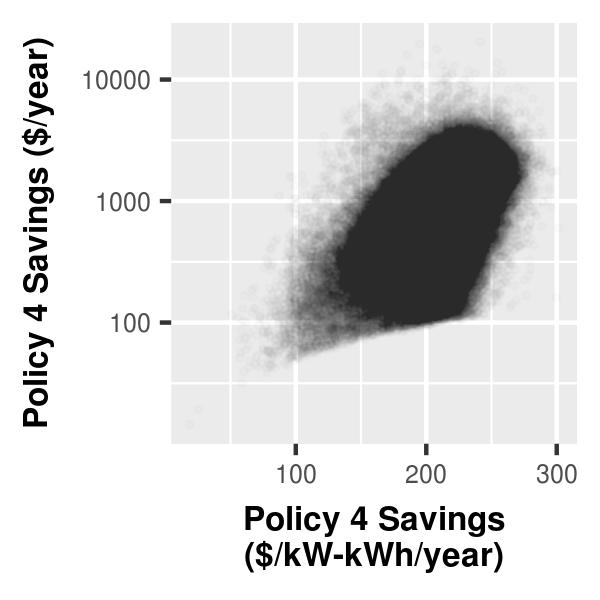}
             \caption{}  
             \label{fig:p4_abs_v_norm}
         \end{subfigure}
        
        \caption{(a) Normalized savings are plotted for Policy 1 (red), Policy 2 (blue), Policy 3 (green), and Policy 4 (brown). For each pricing policy, the households are ranked in decreasing order based on savings. Pricing policies alter both the magnitude and dispersion of savings. (b) Normalized savings under different policies are plotted against each other, with Spearman's rank correlation coefficient $r_s$ given on each plot. Policy 1 and Policy 4 lead to very similar orderings across all households. Other policy comparisons show positive correlations in ordering as well, though not as strong. Recall that under Policy 2, the purchase price varies by location, leading to isolated clouds of points for geographical groupings of households, several of which have strongly similar orderings under different pricing policies.  (c)-(f) Absolute annual savings are not strongly correlated to normalized savings under any of the pricing policies, though there is moderate correlation under Policies 1 and 4. Note that the y-axis is in log scale for these plots.} 
        \label{fig:Savings_and_pairs}
    \end{figure}

Figure \ref{fig:Savings_and_pairs}(a) illustrates the distribution of normalized savings under the different pricing policies.
The variation in normalized savings between households is greatest under Policies 1 and 4.
In the case of Policies 1 and 4, while the purchase price is the same for all households, the sale price is much less than the purchase price, so the benefit from offsetting consumption is much greater than that from selling electricity back to the grid.
Under these two policies, there is a strong correlation between a household's normalized savings and the degree to which its consumption pattern is aligned with the solar irradiance in its area.
That is, heterogeneity in consumption patterns leads to differences in normalized savings.
In contrast, under Policies 2 and 3, the sale price is close to the purchase price.
Households do about as well selling surplus generated electricity back to the grid as they do consuming it themselves, so variations in household consumption patterns do not end up creating large variations in normalized savings.

\subsection{Policy impact on savings}

Policy has a fundamental impact on the magnitude of savings available to the households.
Policies 2 and 3 are the most advantageous to them, allowing households to sell back electricity at almost the full purchase price.
The lowest normalized savings under Policies 2 and 3 are close to the highest under Policies 1 and 4.
Households fare better under Policy 2 than Policy 3 because the dynamic rate is in effect a more pronounced version of the TOU, with higher peaks and lower troughs.
Thus, under Policy 2 PV is worth more because higher prices coincide with peak generation hours, and the storage devices also have higher inter-hour price differentials to work upon.
Switching to Policy 1, which reduces the sale price to the wholesale rate, leads to about a \$100/kW-kWh/year drop in normalized savings for most households as compared to Policy 3.

Interestingly, even though Policy 4 is structured to be unfavorable for rooftop PV, the normalized savings under it are pretty close to those under Policy 1.
The two policies have in common that the sale price is much lower than the purchase price.
Furthermore, both Policy 1 and 4 apply the peak purchase price during evening hours, when household consumption tends to peak.
The storage device allows households to use surplus energy from the sunny hours to offset consumption during evening hours.
Even though the overall structure of their peak and off-peak hours is rather different, the aforementioned factors make make them rather similar when it comes to the valuation of rooftop PV and storage.

\subsection{Policy impact on household savings rank}

Figures \ref{fig:Savings_and_pairs}(b) compares normalized savings under different pricing policies.
Policy impacts which households do best, except in the case of Policy 1 vs. Policy 4, in which case the ordering of households is largely preserved.
The ordering of households by absolute annual savings is almost the same under all pricing policies because the households who consume more have larger net-zero systems and will therefore always do better in absolute terms.
However, the correlation between absolute annual savings and normalized savings is modest at best under all policies, as seen in Figure \ref{fig:Savings_and_pairs}(c) through (f), so households who have larger systems do not always do better by the normalized savings metric.

\section{Value of information}

When deciding how to operate its storage device, a household relies on a forecast of its future consumption and rooftop PV generation.
The preceding sections have assumed that households have perfect foresight of these quantities.
Here we consider the impact of an imperfect forecast, which leads a household to operate its storage suboptimally and therefore to save less money.
We characterize the value of information for a household by evaluating how much more it pays for electricity as the forecast error level increases.
This metric gives a sense of how much a household with PV and storage should be willing to pay a service provider for reducing forecast error through data analytics or improved algorithms.

For each household, we increase the coefficient of variation\footnote{The coefficient of variation is defined as the standard deviation of the forecast residuals divided by the mean of the actuals.} (CV) of its forecast error from 0\% (perfect foresight) to 100\%, and we compute how much its cost of electricity increases in absolute terms given Policy 1 pricing.
The relationship between cost and CV is highly linear in this range.
Thus, the slope of the best fit line between cost and CV provides a good estimate of the sensitivity of the household's electricity cost to forecast error.
We divide the slope by the household net-zero system size to arrive at the value of information for a given household.

\begin{figure}
\centering
  \includegraphics[width=.5\linewidth]{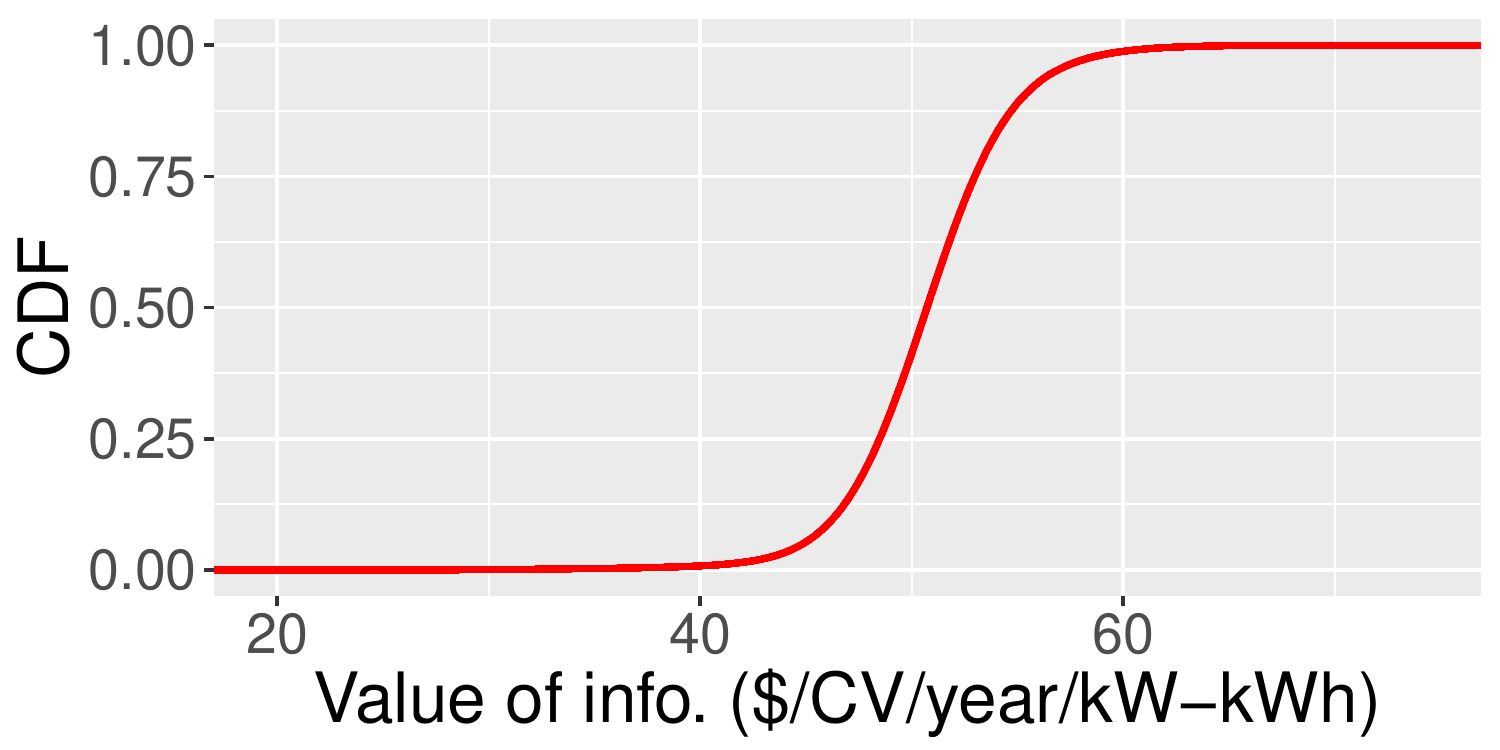}
\caption{The curve gives the distribution of the value of information for households. The value of information is the increase in electricity costs incurred by a household due to an increase in the CV of its forecast errors. These values were computed under Policy 1 pricing.}
\label{fig:Value_of_Info_Distribution}
\end{figure}

As shown in Figure \ref{fig:Value_of_Info_Distribution}, most households have a value of information of \$40-\$60/CV/year/kW-kWh.
Many households face load forecasting error CVs of around 20\%-50\% using well known forecasting methods \cite{sevlian2018}.
Suppose the solar generation forecast errors are similar.
We can estimate the value of halving these forecast errors - which would be no small feat - as approximately \$4-\$15/kW-kWh/year.
This is a price target for information service providers.
This range is small compared to the normalized savings of \$150-\$300/kW-kWh/year which most households achieve under Policy 1 with perfect foresight.
The forecast errors only affect how a household operates its storage device, which plays a relatively small role in generating savings for the household compared to the rooftop PV.

\section{Value of coordination}

We define a coordinator as an entity that collectively manages the storage and rooftop PV of a group of households. Coordinators (or aggregators) can provide services to consumers by enabling the sharing of existing assets. They can also provide services to the grid ecosystem by limiting ramping events caused by synchronized activity, shaving peak load, and regulating voltage and frequency \cite{li2010,kim2017}.

We focus on the additional value that coordination can bring to a group of households that includes adopters and non-adopters.
We define the value of coordination as the additional savings that the coordinator can achieve for the group of households, beyond what they can achieve acting separately.
These additional savings represent a fund that the coordinator can use to compensate the adopters for allowing it to use their devices for the benefit of others in the group.
The value of coordination is the sum of the value of coordinated action (VCA) and the value of coordinated information (VCI), as explained in Figure \ref{fig:Value_of_Coordination_Multi}(a).

The value of coordination depends on the adoption rate and pattern.
The adoption rate is what percentage of households have adopted a rooftop PV and storage system.
The pattern in which households adopt technology is subject to complex economic, behavioral, and social factors.
We consider two simple adoption rules that generate conceptual bounds for adoption possibilities.
Under forward adoption, the adopters are those who stand to save the most in normalized savings terms.
Under reverse adoption, the adopters are those who stand to save the least.
We also consider random adoption, in which households adopt in a random order.

\subsection{Value of coordinated action}

The coordinator can use the capacity of a storage device in one home in order to shift energy usage in other homes.
The coordinator can also ensure that electricity generated by the rooftop PV of homes in the group is used to offset consumption within the group rather than exported to the grid.
In essence, the coordinator manages the group of households as one large aggregated household with a large PV system and storage device, redirecting technology capacity to the uses that decrease the group's total cost the most.
We define the VCA as the additional savings that the coordinator achieves for the group of households beyond what they save when they act separately, in the perfect foresight case.

\subsubsection{A simplified model}

Consider a setting with one time period.
There are two types of households, $A$ and $B$, and $p_A$ is the proportion of type $A$ in the population.
The households all have the same sized solar panels, and they experience the same irradiance, so each household that adopts generates the same energy $e>1$.
Type $A$ households consume load $l_A=e-1$, and $B$ households consume $l_B=e+1$.
The households purchase electricity at price $q$, and they sell surplus back to the utility at $r$, with $q>r$.

If an $A$ household does not have the technology, it pays $q(e-1)$ for electricity.
If it adopts, it pays $-r$ for electricity.
A type $B$ household pays $q(e+1)$ without technology, and $q$ with technology.
Let there be $N$ households in total, and let $f$ be the fraction of households that have adopted a solar panel, assuming the random adoption pattern.
Acting as individuals, the households in total pay:
\begin{subequations}
\begin{align}
C_S&=(1-f)Np_A(e-1)q+(1-f)N(1-p_A)(e+1)q+fNp_A(-r)+fN(1-p_A)q\\
&=(1-f)Nq(e+1-2p_A)+fN((1-p_A)q-p_Ar)
\end{align}
\label{eqn:TC_S}
\end{subequations}

Suppose $p_A>\frac{1}{2}$. When $f\leq f^{\star}\equiv \frac{1+e-2p_A}{e}$, the total energy generated by adopters does not exceed the total load of all households. 
In this case, the total cost paid by the households with coordination is:
\begin{subequations}
\begin{align}
C_C&=(Np_A(e-1)+N(1-p_A)(e+1)-fNe)q\\
&=(1+e-2p_A-fe)Nq.
\end{align}
\end{subequations}
When $f>f^{\star}$, the total energy generated by the adopters exceeds the total load of all households, so now the coordinated group is selling surplus back to the grid. The total cost paid by the households with coordination is now:
\begin{subequations}
\begin{align}
C_C&=(Np_A(e-1)+N(1-p_A)(e+1)-fNe)r\\
&=(1+e-2p_A-fe)Nr.
\end{align}
\end{subequations}
The value of coordinated action is $C_S-C_C$, so:
\begin{equation}
VCA(f)=
    \begin{cases}
      fNp_A(q-r), & \text{if}\ 0\leq f\leq f^{\star}\\
      N(1+e-2p_A-f(e-p_A))(q-r), & \text{if}\ f^{\star}<f\leq 1
    \end{cases}
\end{equation}
The baseline cost of the households is their total bill in the absence of technology, which we denote as $T_{BL}=Nq(1+e-2p_A)$.
We use the baseline cost as a normalizing quantity and write:
\begin{equation}
\frac{VCA(f)}{T_{BL}}=
    \begin{cases}
      \frac{fp_A}{1+e-2p_A}(1-\frac{r}{q}), & \text{if}\ 0\leq f\leq f^{\star}\\
      (1-\frac{f(e-p_A)}{1+e-2p_A})(1-\frac{r}{q}), & \text{if}\ f^{\star}<f\leq 1
    \end{cases}
    \label{eqn:VCA_frxn}
\end{equation}

The VCA is zero when $f=0$. It is linearly increasing in $f$ up to $f^{\star}$ after which it is linearly decreasing, and it remains positive at $f=1$.
This simplified model is a useful explanatory tool that exhibits a rising and falling value of coordinated action that is still positive at 100\% adoption, replicating the major dynamics seen in the actual data using our full multi-period simulation that includes storage.
We proceed to present these results.

\subsubsection{VCA characteristics}

The VCA can be substantial - up to 15\% of the original total cost of electricity, as shown in Figure \ref{fig:Value_of_Coord_no_noise}.
For all three adoption patterns, coordination provides increasing value with increasing adoption, up to a point.
Beyond 35-55\% adoption, which corresponds to a little under half of the maximum system capacity being adopted, the value of coordination decreases.
The reason for this decline is as follows.
The TOU rate has an expensive peak period and an inexpensive off-peak period.
When acting alone to minimize its cost of electricity, a household will use its devices to offset its consumption during the peak period first.
If there is spare capacity after that, it will offset its consumption during off-peak times, and then it will sell electricity back to the grid.
The coordinator, on the other hand, is minimizing the entire group's cost of electricity.
Therefore, it will use any capacity available to first offset as much of the peak period consumption across all households as possible.
Thus, under coordination, more of the adopters' home energy technology capacity goes to offsetting peak period consumption, which means greater savings.

When the level of adoption is low, there are many households who do not have the technology and who therefore need help in offsetting their peak period consumption.
The coordinator provides this help.
When the level of adoption is high, more households have the technology to offset their own peak period consumption.
The coordinator has fewer opportunities to redirect technology capacity to offset peak period consumption, and it instead ends up offsetting off-peak period consumption or selling surplus electricity back to the grid, both of which accrue less savings.
Thus, as adoption increases beyond a certain point, the coordinator is left with lower-value opportunities for redirecting capacity.
In our setting, on the basis of the VCA alone, a coordinator would not encourage forward DER adoption beyond 35\% in a group of households because doing so would reduce the value of its services.

\subsubsection{Impact of distribution service pricing}

The difference between the purchase price and sale price of electricity is a fundamental driver of the VCA, as captured by the presence of the ratio $r/q$ in (\ref{eqn:VCA_frxn}).
When that ratio is very low, as in the case of Policy 1, the retail utility is in essence asserting that distribution services are expensive as compared to generation.
This assertion gives solid ground for coordination arrangements because when a household shares its excess solar generation with its neighbor, that distribution path is much shorter than the distribution path from grid-scale generation that the utility's tariff assumes.
Therefore, the coordinated group should indeed get credited at the full retail rate (or almost the full retail rate) for sharing electricity in this way because they saved the utility both the generation cost and most of the distribution cost.

When the sale price is very close to the purchase price, as in Policies 2 and 3, the VCA is very small.
This type of policy asserts that most of the retail cost of electricity is due to generation.
While these policies preclude much gain from coordination, they lead to higher individual household bill savings as seen in Figure \ref{fig:Savings_and_pairs}(a).
Thus, assuming a given purchase price, the retail utility is faced with a trade-off between incentivizing more DER adoption or incentivizing coordination arrangements.\footnote{Utilities that are not governed by revenue decoupling stand to lose an important source of profit with greater DER adoption. Even utilities that are revenue-decoupled may face difficulties in recovering returns on fixed investments when energy sales decline. For example, raising the price per kWh will incentivize more DER adoption and potentially coordination as well.}

\subsubsection{Mitigating inefficient adoption}

Coordination captures the value that would otherwise be lost due to inefficient adoption patterns by ensuring that the technology capacity is redirected to its most efficient application.
For example, under reverse adoption, the households that stand to save the least adopt first.
The coordinator does very well by redirecting their capacity to those in the group who can save more with it.
Thus the VCA is higher under reverse adoption than under forward adoption.
Even adopters with the highest savings may not take full advantage of their DER capacity every day (e.g. when on vacation), so under forward adoption, the coordinator is still able to realize additional value by reallocating at any given moment capacity that would be underutilized by an adopting household acting on its own.
This ability to dynamically reallocate capacity also explains why the coordinator still delivers value when adoption hits 100\%.

\begin{figure}
\centering
\begin{subfigure}{.49\textwidth}
  \centering
  \includegraphics[width=\linewidth]{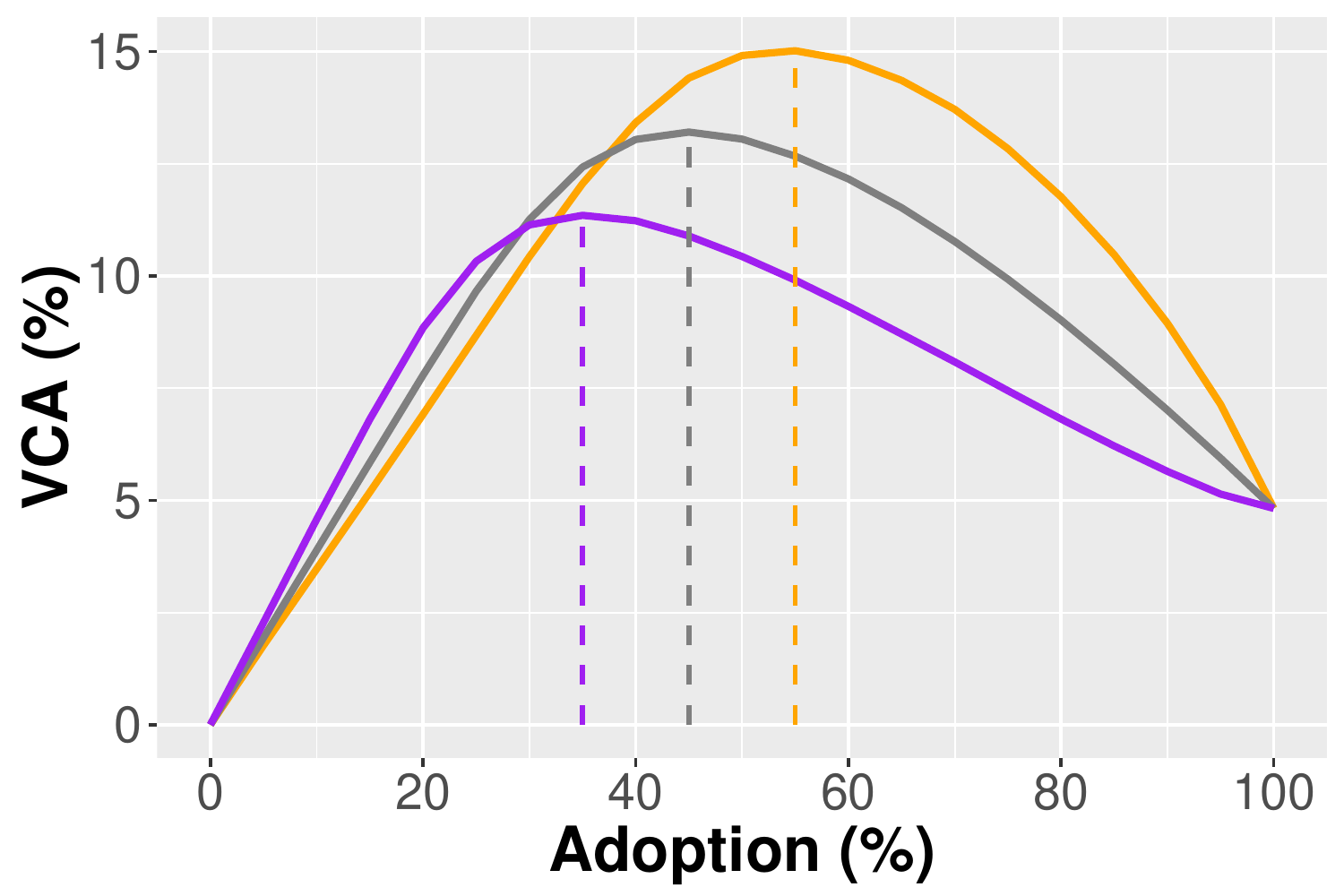}
  \caption{}
  \label{fig:VCA_rate}
\end{subfigure}
\centering
\begin{subfigure}{.49\textwidth}
  \centering
  \includegraphics[width=\linewidth]{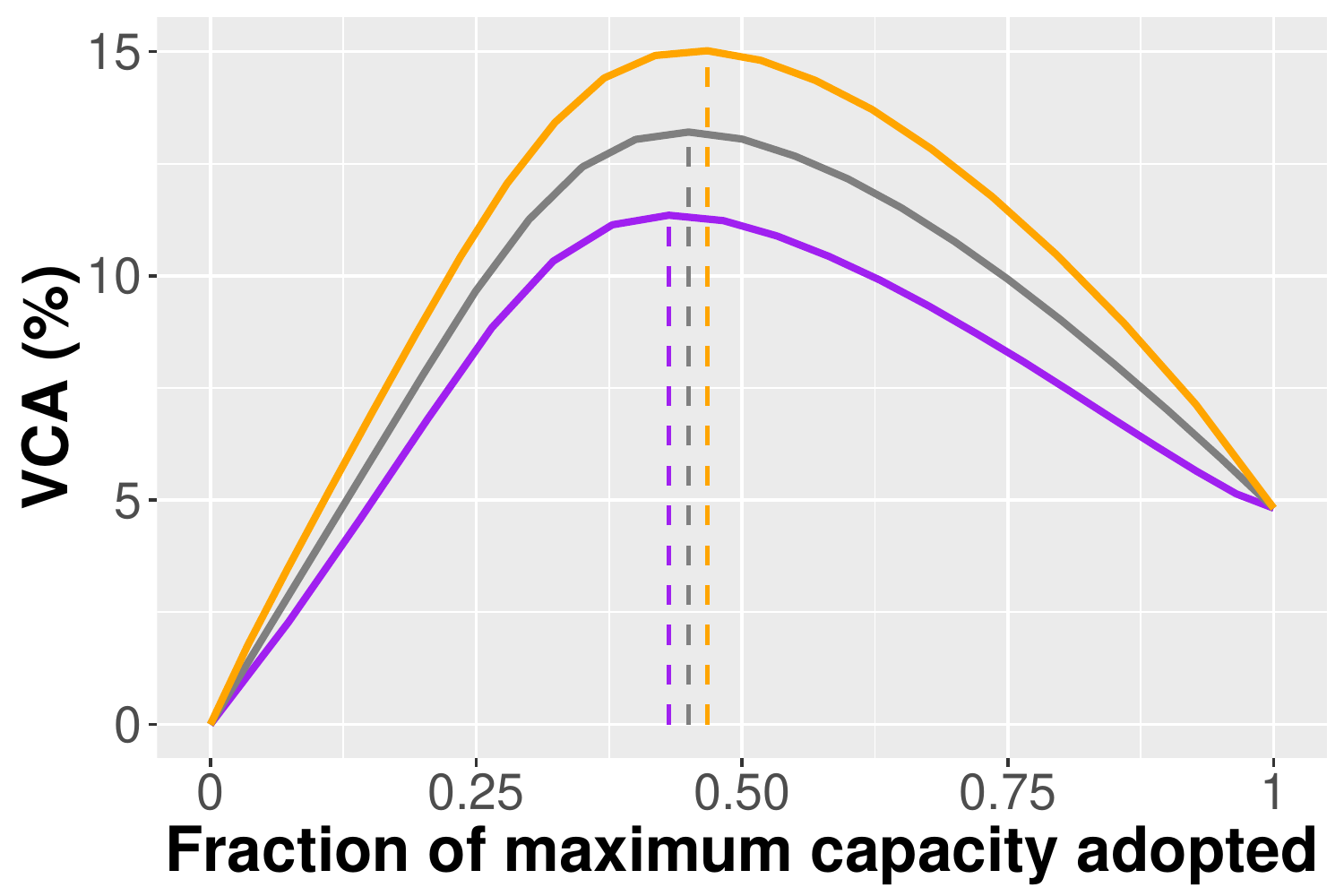}
  \caption{}
  \label{fig:VCA_sys}
\end{subfigure}
\caption{The value of coordinated action (VCA) is given here as a percentage of the baseline total cost of electricity for all households prior to any technology adoption. The purple curve is for the forward adoption pattern, the orange curve is for reverse adoption, and the gray curve is for random adoption. The VCA as a function of household adoption rate is given in (a), and as a function of the fraction of the maximum total capacity adopted is given in (b).
The VCA increases as the level of PV and storage adoption within the group increases up to 35-55\%, or when a little less than half of the maximum system size is adopted.
After that point, coordination provides less value. The VCA is greater under the reverse adoption pattern because the coordinator is able to redirect capacity and therefore overcome inefficient initial allocations of technology. The values here are computed under Policy 1 pricing.
The maximum capacity for the households is just the sum of all of their net-zero system sizes, which in our study is 2.2 GW-GWh.}
\label{fig:Value_of_Coord_no_noise}
\end{figure}

\begin{figure}
\centering
\begin{subfigure}{.3\textwidth}
  \centering
  \includegraphics[width=\linewidth]{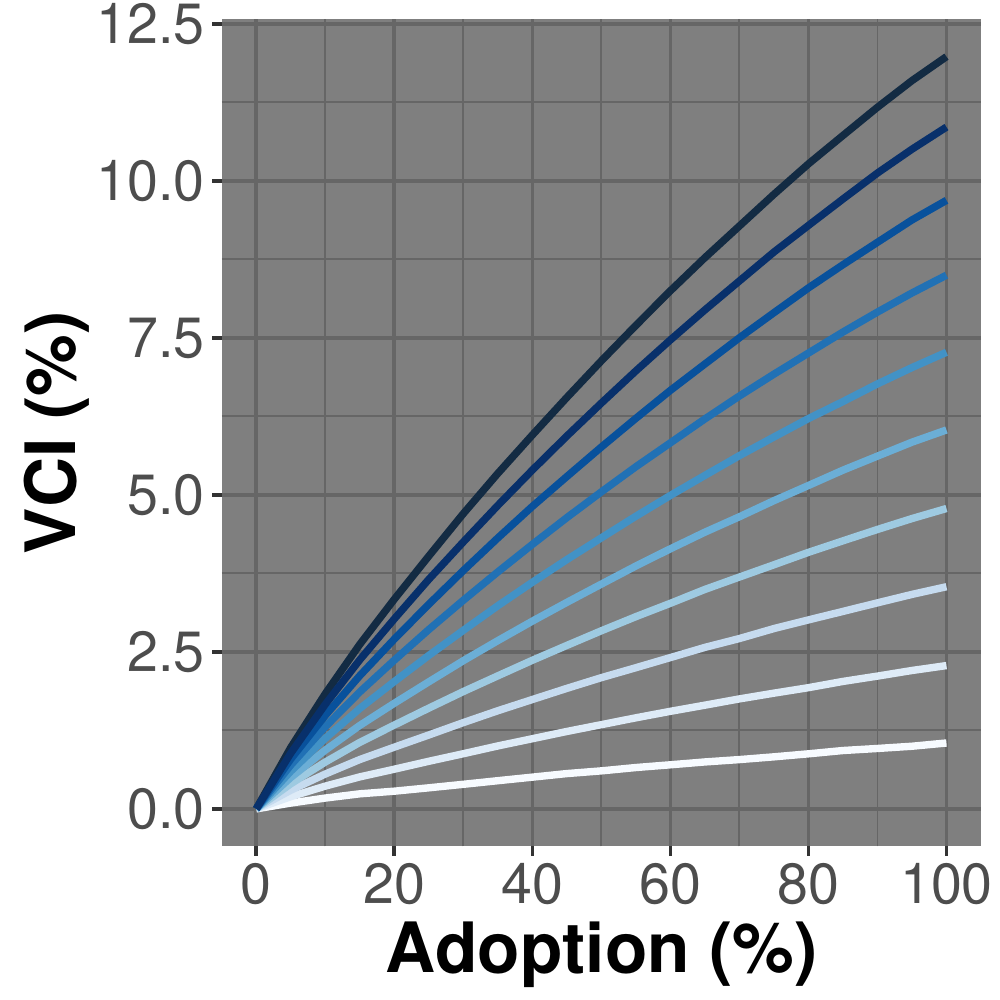}
  \caption{Forward adoption}
  \label{fig:VCI_fwd}
\end{subfigure}
\begin{subfigure}{.3\textwidth}
  \centering
  \includegraphics[width=\linewidth]{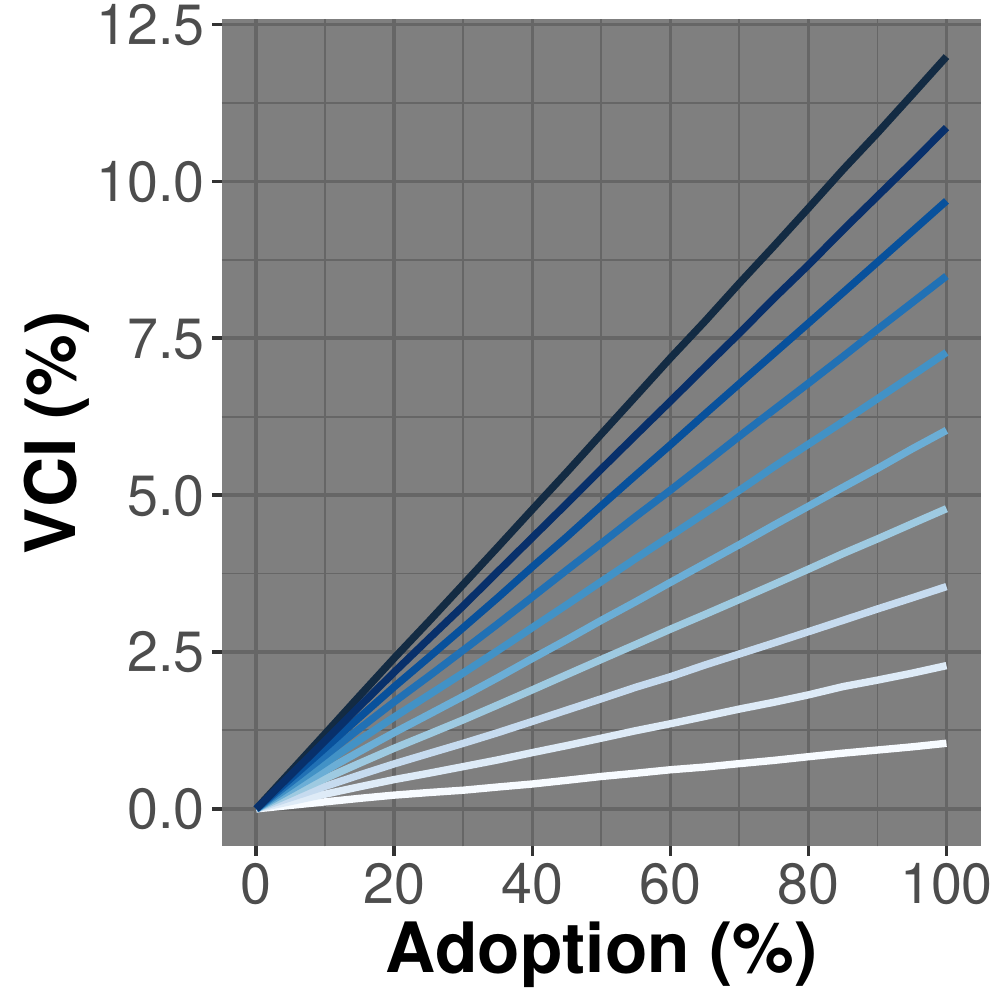}
  \caption{Random adoption}
  \label{fig:VCI_rand}
\end{subfigure}
\begin{subfigure}{.3\textwidth}
  \centering
  \includegraphics[width=\linewidth]{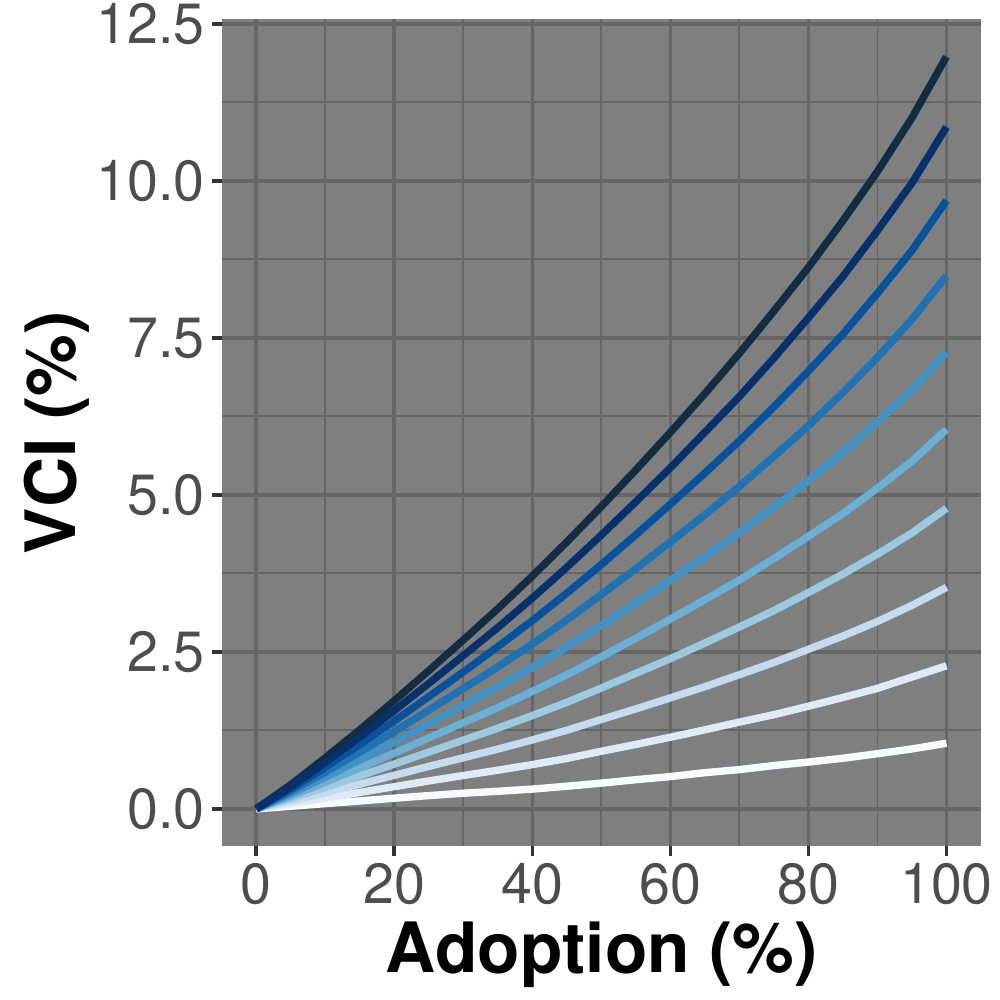}
  \caption{Reverse adoption}
  \label{fig:VCI_rev}
\end{subfigure}
\caption{These plots show the value of coordinated information (VCI) as a function of adoption level and forecast error level for the (a) forward, (b) random, and (c) reverse adoption patterns. In each graph, the lightest line corresponds to a forecast error CV of 10\%, and the darkest a CV of 100\%, with the intermediate colors at 10\% increments. The VCI is greater when the forecast error is greater and when the adoption level is greater. The VCI is reported as a percent of the baseline total electricity cost for all households (the same normalizing quantity used for Fig. \ref{fig:Value_of_Coord_no_noise}). The values here are under Policy 1 pricing.}
\label{fig:VCI}
\end{figure}

\subsection{Value of coordinated information}

\begin{figure}
\centering
\begin{subfigure}[b]{.2\textwidth}
  \centering
  \includegraphics[width=\linewidth]{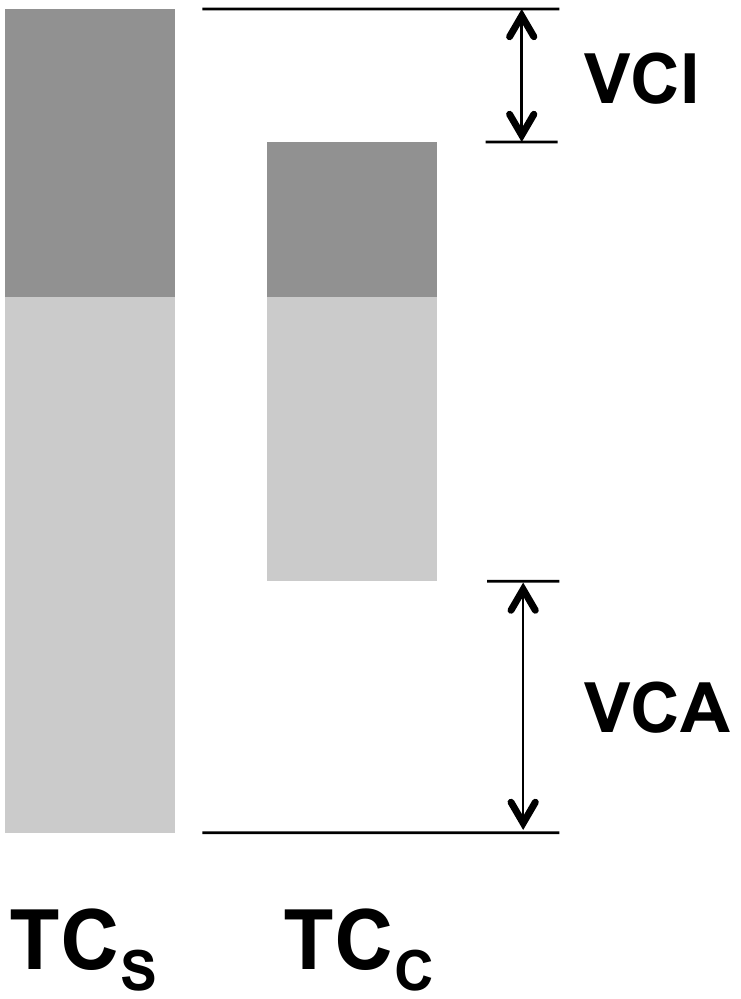}
  \caption{}
  \label{fig:VC_diagram}
\end{subfigure}
\begin{subfigure}[b]{.375\textwidth}
  \centering
  \includegraphics[width=\linewidth]{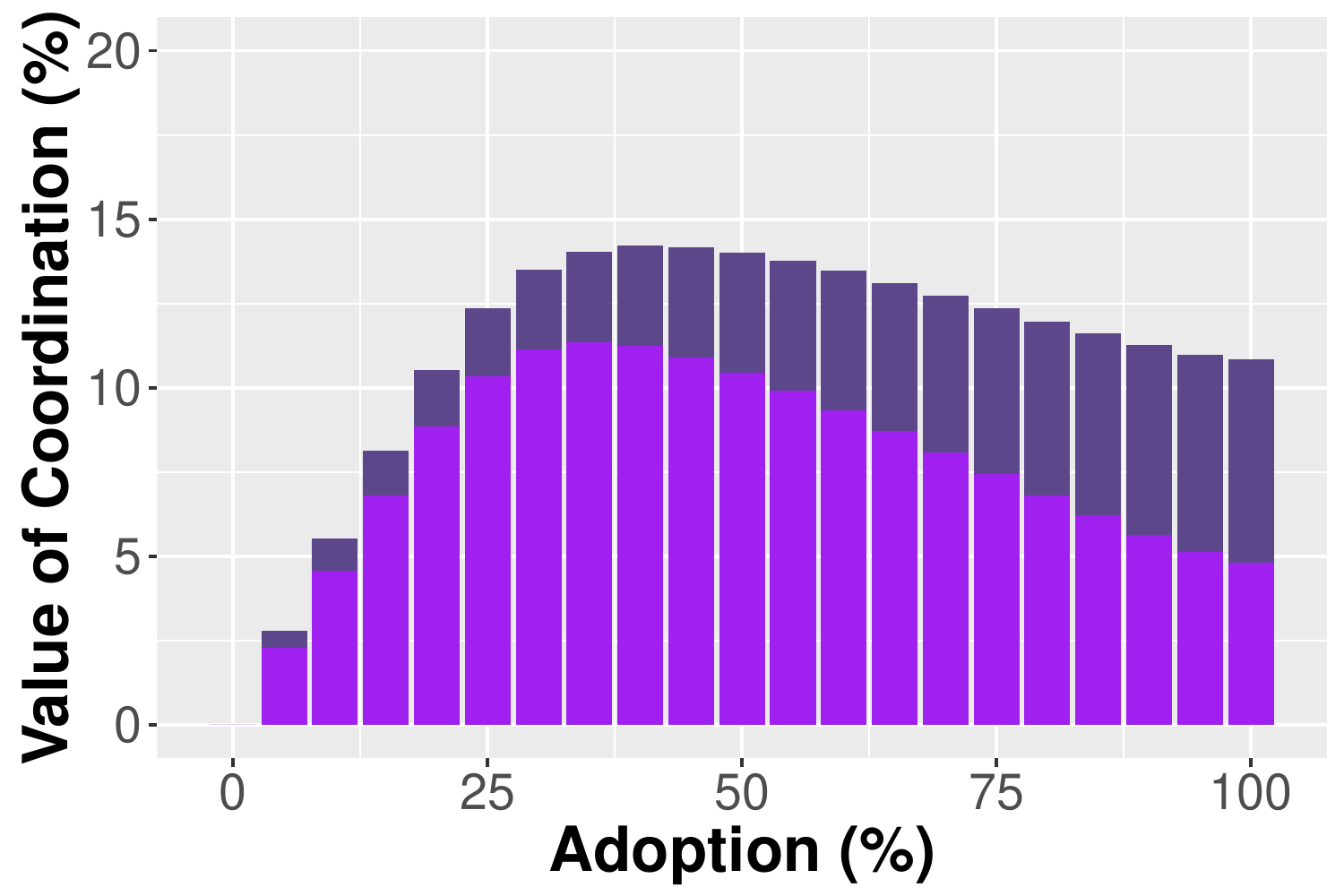}
  \caption{Forward adoption}
  \label{fig:VC_fwd_50}
\end{subfigure}
\begin{subfigure}[b]{.375\textwidth}
  \centering
  \includegraphics[width=\linewidth]{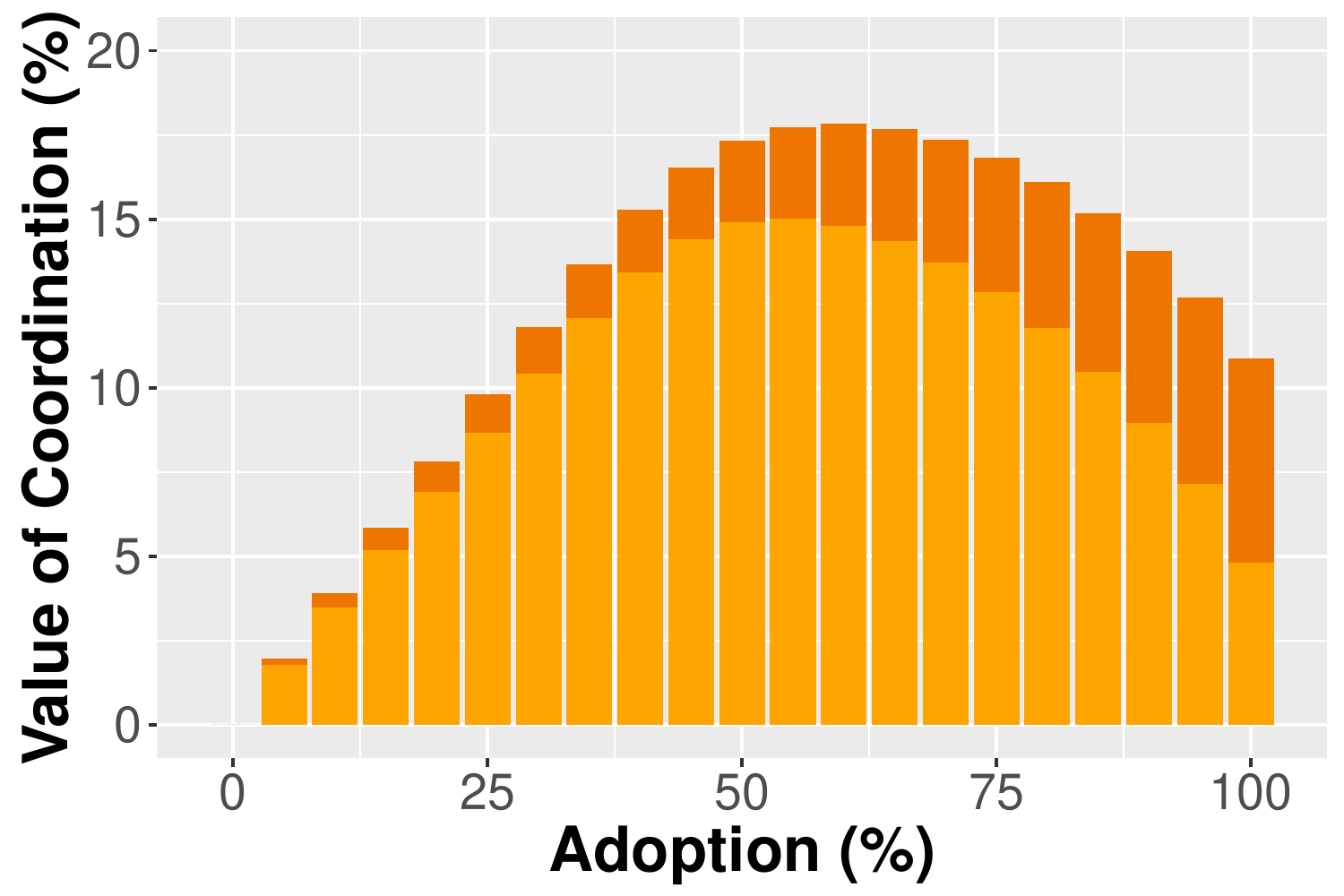}
  \caption{Reverse adoption}
  \label{fig:VC_rev_50}
\end{subfigure}
\caption{(a) This diagram illustrates the definitions of the VCA and the VCI. TC$_{\text{S}}$ is the total cost of electricity paid by the group of households when they optimize separately. TC$_{\text{C}}$ is the total cost they pay under coordination. The light gray bars are costs under perfect foresight, and the dark gray bars are the additional costs due to forecast error. The VCA is the reduction in cost under perfect foresight. The VCI is the additional reduction in cost in the presence of forecast errors. The sum of the VCA and the VCI is the value of coordination. The value of coordination when the forecast error CV is 50\% is plotted for the (b) forward and (c) reverse adoption patterns. The lighter color is the VCA, and the darker color is the VCI. As in Fig. \ref{fig:Value_of_Coord_no_noise}, the values are reported as a percent of the baseline total cost of electricity without technology, and  Policy 1 prices apply.}
\label{fig:Value_of_Coordination_Multi}
\end{figure}

The VCA is defined under perfect foresight.
The coordinator delivers an additional benefit in the presence of uncertainty.
When operating the aggregated storage devices of the adopters, the coordinator only needs to forecast aggregate quantities.
Thus it faces much lower forecast errors than the individual households \cite{sevlian2018}, enabling it to achieve a more optimal outcome.
We define the value of coordinated information as the additional savings beyond the VCA that the coordinator achieves in the presence of forecast error.
The VCI increases with adoption level and with forecast error CV for all adoption patterns, as shown in Figure \ref{fig:VCI}.
The adoption pattern has a small effect on the VCI, with the VCI being greater under forward adoption.
At higher levels of adoption, the VCI increases more for a given increase in forecast error level.
This is because when adoption is higher, the coordinator is operating a greater amount of technology capacity, so its forecasting advantage yields a greater amount of additional savings.
Figures \ref{fig:Value_of_Coordination_Multi}(b) and (c) show the value of coordination as the sum of the VCI and VCA for the forward and reverse adoption patterns at a forecast error level of 50\%.
For both adoption patterns, at higher levels of adoption the increasing VCI is more than offset by the drop in the VCA, so a coordinator would not encourage adoption beyond a certain point, about 40\% for forward adoption and 55\% for reverse adoption.
In general, the value of coordination may be increasing or decreasing in adoption rate depending on the adoption pattern, adoption level, forecast error level, and adopted system size.

\section{Discussion}

Our methodology and results are useful to stakeholders developing their plans of action with respect to behind-the-meter DERs.
This study provides policy makers with high-level guidance about the impact of different pricing policies on household incentives and ordering.
Pricing policy is a useful lever to adjust the magnitude of normalized savings available to households and hence their economic incentives to adopt DERs.
Pricing policy significantly impacts which households save the most, as well as the dispersion of savings - how wide the gap is between the top and the bottom saver.
The one unexpected exception is that savings are very similar under Policies 1 and 4 - even though they have very different peak hour timing, and even though Policy 4 does not compensate households at all for energy supplied back to the grid.

Enabling sharing arrangements is another way that policy makers can ensure that as many households as possible can benefit from DERs.
Normalized savings are not strongly correlated with absolute savings.
That means that there are many households whose net-zero system size may be small but who would derive relatively large per-unit value from those systems and could therefore afford relatively large per-unit system costs.
Devising policy arrangements and business models for community or neighborhood pooling for solar and storage systems would allow these households to overcome barriers associated with small scale, such as fixed costs and commercial availability.
Related proposals such as shared solar and virtual net-metering are under study \cite{feldman2015}.

The magnitude and distribution of household savings gives vendors of DER equipment a sense of their market size and structure.
These vendors sell equipment with costs that vary with system size, so the normalized saving metric is the most relevant.
Under Policy 3, most households save about the same amount on a per-unit basis.
Thus, vendors would do equally well to pool together households with small net-zero system sizes or to pursue individually households with large net-zero system sizes.
By contrast, under Policy 1, there is a modest positive correlation between normalized and absolute savings.
The vendor would do best by targeting households with larger net-zero system sizes, avoiding the complexities of pooling together households with smaller loads who cannot afford to pay as much on a per-unit basis.

The value of information analysis gives guidance to information service providers about potential price points for algorithms and analytics that help households manage their storage devices.
Similarly, the value of coordination analysis gives aggregators a sense of the revenue possibilities for sharing DERs among households.
We identify a tipping point for these aggregators.
When the adoption rate within a group of households exceeds a certain level, the VCA declines.
In the presence of uncertainty, however, the aggregator can deliver a VCI that increases in adoption rate.
The combined value of coordination may increase or decrease with increasing adoption, depending on the adoption pattern, adoption level, and uncertainty level.
Aggregators need to take these factors into account when determining how to assemble households into groups of adopters and non-adopters in a mix that allows the aggregator to deliver maximum value.

With coordination, even if the households with the strongest economic incentives are the ones who adopt DERs, other households can share in their capacity and reduce their costs as well.
All households are better off with coordination, and the technology capacity is deployed to its most efficient applications.
The regulatory and policy environment for this type of coordination will need to emerge.
Large apartment buildings are a natural first case for coordination that remains within a property boundary.
Beyond that, policy that enables coordination will have to take into account the technical and economic impacts on the distribution grid of sharing solar and storage resources through it.
Such considerations may limit the size of groups that aggregators can assemble and coordinate.
With smaller groups, the VCA will still scale as in Figure \ref{fig:Value_of_Coord_no_noise}, but the VCI may be smaller because the scaling law for reducing forecast error is dependent on group size.

Retail utilities will need to price distribution services in a way that reflects the impact of distributed energy resources on lowering distribution costs.
If households are consuming more of their electrical energy from more local sources, their costs of distribution should decrease.
Pricing policies that overvalue distribution services create strong grounds for coordination arrangements that allow groups of households to capture even more bill savings from the utility.
Proposals like electricity markets at the distribution grid level, analogous to the transmission-level market operated by independent system operators, are one possible means to achieve correct energy and distribution pricing as DER adoption increases.

Finally, we note a tension between the business models of the equipment vendor and the coordination service provider.
An equipment vendor interested in selling PV and storage systems to households does best by targeting those with the highest potential savings because they can pay the most for the equipment.
It will be inclined to proceed along the forward adoption pattern so that it can charge higher prices and get more customers.
On the other hand, a coordination service provider would want to encourage the reverse adoption pattern because that makes its services more valuable.
However, the coordination service provider will not encourage adoption beyond the point at which the value of its services is maximized.
The tension between these two business models could manifest in support for competing policies dealing with pricing, equipment subsidies, and sharing mechanisms.

\section*{Acknowledgements}

We thank Pacific Gas and Electric Company for providing the smart meter data used in this study.
We thank Dr. Charles Kolstad and Dr. Stefan Reichelstein for providing valuable feedback about the economics of the value of coordinated action.

\section{Methods}

Our study incorporates actual smart meter data, pricing data, solar irradiance data, and specifications for currently available home energy technologies.
We assume that households and groups of households operate the technologies to minimize their cost of electricity.
Here, we describe in detail the data sources we use and the analyses we perform.

\subsection{Data sources}

\subsubsection{Household consumption}

The household electricity consumption data comes from over 500,000 residential smart meters over a one year period spanning from November 2011 to October 2012.
The households are all customers of Pacific Gas and Electric Company (PG\&E) in California.
These meters include single family homes and apartments.
Meters with very low consumption ($<$0.1 kW annual mean) are excluded, as are meters with a high amount of zero readings ($>$50\% of all readings).

We treat the smart meter data that we have for a household as its inflexible load $\textbf{L}_i$.
This move rests on two assumptions.
The first is that consumption behavior does not change due to rate changes.
Thus, even though almost all of these households were on an inclining block rate for the period of time of the meter data, their electricity consumption behavior does not change when exposed to the pricing policies in our study.
Note that retail rate design typically aims for revenue neutrality, which means that between different rates, the marginal price faced by consumers varies much more than the average price.
Both \cite{ito2014} and \cite{borenstein2009} find that household electricity consumption responds most strongly to the average price of electricity, lending support to using the meter data as the inflexible load.

The second assumption is that consumption behavior does not change due to DER adoption.
For the sizes of PV systems we study, \cite{mcallister2012} suggests that most households would have very small changes in their overall electricity consumption.

\subsubsection{Prices}

\paragraph{Retail TOU}

Utilities are moving towards time-varying rates, which create incentives for consumers to shift their electricity purchases to periods when the rate is lower \cite{cpuc2015}. A version of PG\&E's E-TOU Option B serves as the retail TOU rate for this study \cite{pgeetou}.
Table \ref{tab:TOU_rate} contains the relevant elements of the tariff.
We exclude fixed charges in the tariff when computing household electricity bills.

\begin{table}
\centering
\begin{tabular}{cc|c}
\multicolumn{1}{l}{}                  & Off peak hours   & Peak hours (4pm-9pm)                  \\ \cline{2-3} 
\multicolumn{1}{c|}{June - September} & \$ 0.25511 / kWh & \multicolumn{1}{c|}{\$ 0.35817 / kWh} \\ \hline
\multicolumn{1}{c|}{October - May}    & \$ 0.20191 / kWh & \multicolumn{1}{c|}{\$ 0.22071 / kWh} \\ \cline{2-3} 
\end{tabular}
\caption{The retail TOU rate for this study comes from PG\&E's E-TOU. Peak hour rates apply only on non-holiday weekdays. On holidays and weekends, all hours are charged at the off peak rate.}
\label{tab:TOU_rate}
\end{table}

\paragraph{Discounted TOU}

In a recent survey, industry professionals were asked what households should receive for electricity they sell back to the grid.
A plurality of them favored the full retail rate minus the costs of using the physical infrastructure of the grid \cite{gtm2016}.
These costs are estimated as 20\% of the retail rate, so households would be compensated at 80\% of the retail rate - what we call the discounted TOU.

\paragraph{Wholesale}

Locational marginal prices (LMPs) serve as the wholesale price.
They come from a wholesale energy market administered by the California Independent System Operator (CAISO).
We use the day ahead LMPs published by CAISO for the dates corresponding to the household consumption data, but we set any negative prices to zero.

The wholesale rate varies by location.
We use a rough mapping from a household's zip code to the nearest CAISO pricing node by latitude and longitude, and we use the LMP from that node as the wholesale price for the household.
Thus all households in a given zip code face the same wholesale price.

\paragraph{Retail dynamic}

The retail dynamic rate is a scaled version of the wholesale rate.
The scaling is designed to maintain day-by-day revenue neutrality with the retail TOU.
For each day, across all zip codes, the total cost of all of the households' inflexible load at the retail TOU rate and the wholesale rate is computed.
The LMPs for each zip code are then scaled up by the ratio of the retail TOU revenue to the wholesale revenue.
The retail dynamic rate exposes households more directly to the fluctuations of the wholesale market.

\paragraph{Discounted dynamic}

The discounted dynamic rate is 80\% of the dynamic rate.

\paragraph{Flipped TOU}

We designed the flipped TOU to be a speculative worst-case time-of-use rate in terms of disincentivizing PV adoption.
We had three goals: the off-peak (lower) price should apply during the hours of maximum solar irradiance, i.e. 9am-3pm; summer should be cheaper than winter; and the gap between peak hour and off-peak hour prices should be similar to the gap in the Retail TOU rate.
Furthermore, we enforced revenue neutrality with the Retail TOU rate.
We came up with the following rate: 30 \textcent/kWh during winter peak hours; 20 \textcent/kWh during winter off-peak hours; 25 \textcent/kWh during summer peak hours; 15 \textcent/kWh during summer off-peak hours.

\subsubsection{Solar irradiance}
\label{Sec:Irradiance}

The only geographic information we have for the households is their zip code.
We have a latitude and longitude pair for each zip code, the same as was used for mapping the wholesale price.
We use the global horizontal irradiance (GHI) for this latitude and longitude from version 3 of the Physical Solar Model of the National Solar Radiation Database \cite{habte2017}, for the same time period as in the household consumption data.
All households in a given zip code are assigned the same solar irradiance.

\subsubsection{Home technologies}

We assume that households can install DERs behind the meter without interference from their electricity provider, and that households living in apartments can adopt DERs by participating in a building or neighborhood sharing arrangement \cite{feldman2015,funkhouser2015}.

\paragraph{Interconnection}

We assume that the rooftop PV is connected to the house AC bus through an inverter with efficiency $\eta_I=92\%$.
The storage device is connected in the same way.
Thus, all energy flows to and from the storage device, and from the PV, incur 8\% loss through the inverters.
We assume that there is no limit on the power that a household can draw from or provide to the grid.

\paragraph{Rooftop PV}

We size the rooftop PV system to make the household net-zero electrical energy as follows.
Let $\textbf{V}_i$ be the irradiance sample for a household as described in Section \ref{Sec:Irradiance}. Then $z_i=(\textbf{1}^T\textbf{L}_i)/(\eta_I\textbf{1}^T\textbf{V}_i)/\textrm{m}^2$.
\footnote{The load has units of kWh, and the irradiance has units of kWh/$\textrm{m}^2$. Their ratio has units of $\textrm{m}^2$, which would be of importance if we were trying to compute panel areas using conversion efficiencies. Given that we are not interested in panel areas, we divide further by $\textrm{m}^2$ to obtain a dimensionless $z_i$. If one assumes the PV system converts irradiance to electrical energy at an efficiency of $\zeta$, then the net-zero panel would need to have an area of $(z_i/\zeta)\,\textrm{m}^2$}
The distribution of $z_i$ is given in Figure \ref{fig:Diagram_and_Scenarios}(c).
As a reference point, according to the Solar Energy Industries Association, 5 kW was the average size of installed residential systems in the United States a few years ago \cite{seia}.

\paragraph{Storage device}

The storage device specifications are scaled to the second generation of the Tesla Powerwall, which has a usable capacity of 13.5 kWh, a maximum sustained charging rate of 5 kW, and a maximum sustained discharging rate of 5 kW.
We set the household's storage device to have a capacity of $z_i$ kWh, i.e. it has the same capacity in kWh as the kW rating of its net-zero PV system.
The household storage device's maximum charging and discharging rates are then $z_i\cdot\frac{5}{13.5}\,\textrm{kW}$.

We set the round trip efficiency to 0.92, and we take the square root of that, 0.959, to split it into a charging and discharging efficiency.
We assume a 5\% daily rate of self-discharge, or a 95\% daily self-discharge efficiency, which we convert into a 99.8\% hourly self-discharge efficiency by taking its 24th root.

\subsection{Pricing scenarios}

We analyze four different pricing policies, as listed in the table in Figure \ref{fig:Diagram_and_Scenarios} of the main text.
Under Policy 1, households purchase electricity at the retail TOU rate, and they sell any hourly surplus back to the grid at the wholesale rate.
This is generally considered the least generous compensation scheme that households are likely to encounter.
Note that this policy is much less favorable to PV than typical net-metering arrangements, which allow households to rollover excess generation from day to day and month to month, with a financial settlement of any surplus taking place at the end of the year.
In our setup, the financial settlement of net shortage or surplus occurs hourly for all four policies.
Under Policy 2, households purchase at the retail dynamic rate and sell surplus back at the discounted dynamic rate.
Under Policy 3, households purchase electricity at the retail TOU, and they sell it back to the grid at the discounted retail TOU.
Under Policy 4, households purchase electricity at the flipped TOU, and they get no compensation for supply electricity back to the grid (i.e. a sale price of zero).
The retail TOU, discounted retail TOU, and flipped TOU rate are the same for all households.
Thus, all households face the same prices under Policies 3 and 4.

In all four scenarios, electricity prices are exogenous\footnote{Conventional residential retail rates do not change rapidly, so for all intents and purposes the retail TOU rate in our study would be fixed as households begin the uptake of DERs.}; there is no dynamic market clearing in our study.
In other words, households are price takers; their actions do not influence the prices they face.

We require that the sale price be less than or equal to the purchase price at each point in time to preclude instantaneous and unlimited arbitrage.
Thus, under Policy 1, if the wholesale rate is greater than the TOU rate in a given hour, the household only receives the TOU rate for any excess electricity it sells back to the grid.

\subsection{Operation of storage device}

We consider rooftop PV an uncontrollable system - it will produce however much power is dictated by the solar irradiance.
The storage device, however, is controllable, and households have to determine how to operate it.
In our study, households solve a linear program on a daily basis to choose an optimal schedule for charging and discharging the battery.
Here we describe that linear program.

$\textbf{L}_i$ is the $i$th household's hourly smart meter data for a year, and $\textbf{l}_i^{(j)}$ is the data for the $j$th day, so that $\textbf{L}_i=[\textbf{l}_i^{(1)}\cdots\textbf{l}_i^{(Y)}]$, where $Y$ is the number of days in the year (366 in our case). We take $\textbf{l}_i^{(j)}$ as the household's inflexible electrical energy consumption for the day.
Similarly, $\textbf{V}_i$ is the household's hourly solar irradiance for a year, and $\textbf{v}_i^{(j)}$ is the irradiance for the $j$th day.
Let the household's hourly PV energy generation be $\textbf{E}_i=z_i\textbf{V}_i$, with $\textbf{e}_i^{(j)}$ being the energy generated by the PV system on the $j$th day.
(If the household doesn't have a PV system, then $\textbf{e}_i^{(j)}=\textbf{0}$.)
Finally, we get the household's net load for the day, $\textbf{n}=\textbf{l}_i^{(j)}-\eta_I\textbf{e}_i^{(j)}$, where $\eta_I$ is the inverter efficiency.
In this section we assume that households have perfect foresight of $\textbf{l}_i^{(j)}$ and $\textbf{e}_i^{(j)}$ for the coming 24 hours.

Let $\textbf{Q}_i$ and $\textbf{R}_i$ be the year-long vectors of hourly prices at which the household buys and sells electricity, respectively, under the given pricing policy.
Let $\textbf{q}_i^{(j)}$ and $\textbf{r}_i^{(j)}$ denote the prices on the $j$th day.

Each day, the household must select  $\textbf{u}^{(j)}$, a sequence of hourly actions for the storage device.
Let $\textbf{u}_h^{(j)}$ denote the action during hour $h$ of the day.
When $\textbf{u}_h^{(j)}$ is positive, the device is charging in hour $h$; when it is negative, the device is discharging.
Similarly, $\textbf{x}$ denotes the state of charge of the device, and $\textbf{x}_h$ the state of charge at the end of hour $h$.
Let $\eta_C$ and $\eta_D$ be the charging and discharging efficiencies, respectively, and $\eta_S$ be the self-discharge efficiency.
Finally, $\textbf{g}$ denotes the hourly net exchange with the grid.
When $\textbf{g}_h$ is positive, the household is drawing power from the grid during hour $h$; when it is negative, the household is supplying power to the grid.

The household seeks to minimize what it pays for electricity this day.
It solves the following linear program.

\begin{subequations}
\label{eqn:Optimization}
\begin{align}
\underset{\textbf{u}^{(j)},\textbf{x},\textbf{g} \; \in \; \mathbb{R}^{24}}{\text{minimize}} \; & [\textbf{g}]_+^T\textbf{q}_i^{(j)} + [\textbf{g}]_-^T\textbf{r}_i^{(j)} \\
\mbox{subject to} \; & \textbf{g}=\textbf{n}+\frac{1}{\eta_C\eta_I}[\textbf{u}^{(j)}]_++(\eta_D\eta_I)[\textbf{u}^{(j)}]_- \\
& -\underline{u}_i \leq \textbf{u}^{(j)} \leq \overline{u}_i \\
& 0 \leq \textbf{x} \leq \overline{x}_i \\
& \textbf{x}_0 = x_0 \\
& \textbf{x}_h = \eta_S\textbf{x}_{h-1}+\textbf{u}_h^{(j)},\; \forall h \in \lbrace1,\ldots,24\rbrace,
\end{align}
\end{subequations}

where $\overline{u}_i$ and $\underline{u}_i$ are respectively the maximum charge and discharge rates for household $i$'s storage device, $\overline{x}_i$ is its maximum capacity, and $x_0$ is the state of charge of the device at the end of the prior day.
The $[\cdot]_+$ and $[\cdot]_-$ operators represent element-wise application of $\textrm{max}(\cdot,0)$ and $\textrm{min}(\cdot,0)$, respectively.

Let the optimal schedule be denoted $\textbf{u}^{(j)\star}$.
The household operates the battery per $\textbf{u}^{(j)\star}$, so the minimized objective is what it pays for electricity this day.
Denote the minimized objective value as $c_i^{(j)}$.

\subsection{Household savings analysis}

For a given pricing policy, we first compute the baseline annual bill for the household in the absence of the home energy technologies: $b_{BL,i} =\textbf{L}_i^T\textbf{Q}_i$.
Next, we compute the household's new bill in the presence of the technologies for the given pricing policy: $b_{N,i} = \sum_{j=1}^{Y}c_i^{(j)}$.
The absolute annual savings $S_{a,i}=b_{BL,i}-b_{N,i}$.
The normalized savings $S_{n,i} = S_{a,i}/z_i\textrm{ kW-kWh}$.
Figure \ref{fig:Savings_and_pairs}(a) plots the distribution of $S_{n,i}$ under different pricing policies, and Figures \ref{fig:Savings_and_pairs}(c)-(f) compare $S_{n,i}$ and $S_{a,i}$ under different policies.

We compute error bars for the absolute annual savings and normalized savings by bootstrap resampling from the $Y$ days of the year for each household.
The error bars are very small, so we omit them in the graphs.

\subsection{Value of information analysis}

We estimate the value of information as the additional cost incurred by a household due to an increase in forecasting errors.

For a given day, instead of having perfect foresight of its inflexible consumption $\textbf{l}_i^{(j)}$ and generation $\textbf{e}_i^{(j)}$, the household must generate forecasts $\hat{\textbf{l}}_i^{(j)}$ and $\hat{\textbf{e}}_i^{(j)}$.
Note that the household will know all relevant prices without uncertainty because the retail TOU does not change on a daily basis, and the day ahead LMPs are published prior to the start of the day on which they apply.

We introduce forecast error based on the coefficient of variation \cite{sevlian2018}.
Let $\hat{\textbf{l}}_{i,h}^{(j)} = [ \textbf{l}_{i,h}^{(j)}+\epsilon_h^{(j)}]_+$, and $\hat{\textbf{e}}_{i,h}^{(j)} = [ \textbf{e}_{i,h}^{(j)}+\gamma_h^{(j)}]_+$ where $\epsilon_h^{(j)}$ and $\gamma_h^{(j)}$ are distributed independently across $j$ and independently and identically across $h$.
For the forecast error CV of $P$, $\epsilon_h^{(j)}$ is normally distributed with $\mu=0$ and $\sigma=P\bar{\textbf{L}}_i$, where $\bar{\textbf{L}}_i$ is the mean of all of the entries of $\textbf{L}_i$.
In other words, $\bar{\textbf{L}}_i$ is the mean hourly inflexible load of the $i$th household.
Similarly, $\gamma_h^{(j)}$ is normally distributed with $\mu=0$ and $\sigma=P\bar{\textbf{E}}_i$, where $\bar{\textbf{E}}_i$ is the mean hourly solar generation for household $i$.

The household uses these forecasts to schedule the operation of its storage device.
Let $\hat{\textbf{n}}=\hat{\textbf{l}}_i^{(j)}-\eta_I\hat{\textbf{e}}_i^{(j)}$.
The household solves the same linear program as before, except that $\hat{\textbf{n}}$ replaces $\textbf{n}$ in constraint (\ref{eqn:Optimization}b).
The optimal schedule $\textbf{u}^{(j)\dagger}$ from this solution is what the household uses to operate its storage device.

The household's cost of electricity in this case is \emph{not} the objective from the linear program because the true consumption and solar generation are different from the forecasts.
Let $\textbf{g}^{\dagger}=\textbf{n}+\frac{1}{\eta_C\eta_I}[\textbf{u}^{(j)\dagger}]_++(\eta_D\eta_I)[\textbf{u}^{(j)\dagger}]_-$, where $\textbf{n}$ is the true net load, i.e. $\textbf{n}=\textbf{l}_i^{(j)}-\eta_I\textbf{e}_i^{(j)}$.
Then the household's cost of electricity for this day is $c_i^{(j)\dagger}=[\textbf{g}^{\dagger}]_+^T\textbf{q}_i^{(j)} + [\textbf{g}^{\dagger}]_-^T\textbf{r}_i^{(j)}$.

The household's annual cost of electricity at this level of forecast error CV is $b^{\dagger}_{N,i}(P) = \sum_{j=1}^{Y}c_i^{(j)\dagger}$.
We compute $b^{\dagger}_{N,i}(P)$ for $P=10,20,\dots,100$, and $b^{\dagger}_{N,i}(0)=b_{N,i}$.
We then perform a linear regression of $b^{\dagger}_{N,i}(P)$ on $P$.
The value of information for household $i$ is the slope coefficient from this regression divided by the household's net-zero system size $z_i$ kW-kWh.
The cumulative distribution of this metric, which has units of \$/CV/year/kW-kWh, under Policy 1 prices is given in Figure \ref{fig:Value_of_Info_Distribution}.

\subsection{Value of coordination analysis}

For a given scenario, we compute the value of coordination as the additional savings that the entire group of about 500,000 households could obtain if they acted collectively instead of as individuals.
We break the value of coordination into two components: the value of coordinated action (VCA) and the value of coordinated information (VCI).

\subsubsection{Total costs without coordination}

Index all households by $i=1,\ldots,N$, where $N$ is the total number of households.
Under a given pricing policy, let $f(i)$ be the ordering position of  household $i$ when the households are sorted in decreasing order of $S_{n,i}$.
Let $r(i)$ be the ordering position of household $i$ when the households are sorted in increasing order of $S_{n,i}$.
Finally, let $s(i)$ be the ordering position of household $i$ in a random permutation.
Thus, if household $m$ saved the most under this policy, then $f(m)=1$, $r(m)=N$, and $s(m)$ is wherever $m$ is in the random permutation.

Consider a given adoption rate, $t$\%.
Define $A_{\textrm{fwd}}(t) = \lbrace i:f(i) \leq N\cdot t\% \rbrace$, $A_{\textrm{rev}}(t) = \lbrace i:r(i) \leq N\cdot t\% \rbrace$, and $A_{\textrm{rdm}}(t) = \lbrace i:s(i) \leq N\cdot t\% \rbrace$.
Define $B_{\textrm{fwd}}(t) = \lbrace 1,\ldots,N \rbrace \setminus A_{\textrm{fwd}}(t)$, and define $B_{\textrm{rev}}(t)$ and $B_{\textrm{rdm}}(t)$ analogously.

\begin{figure}
\centering
\includegraphics[width=.6\textwidth]{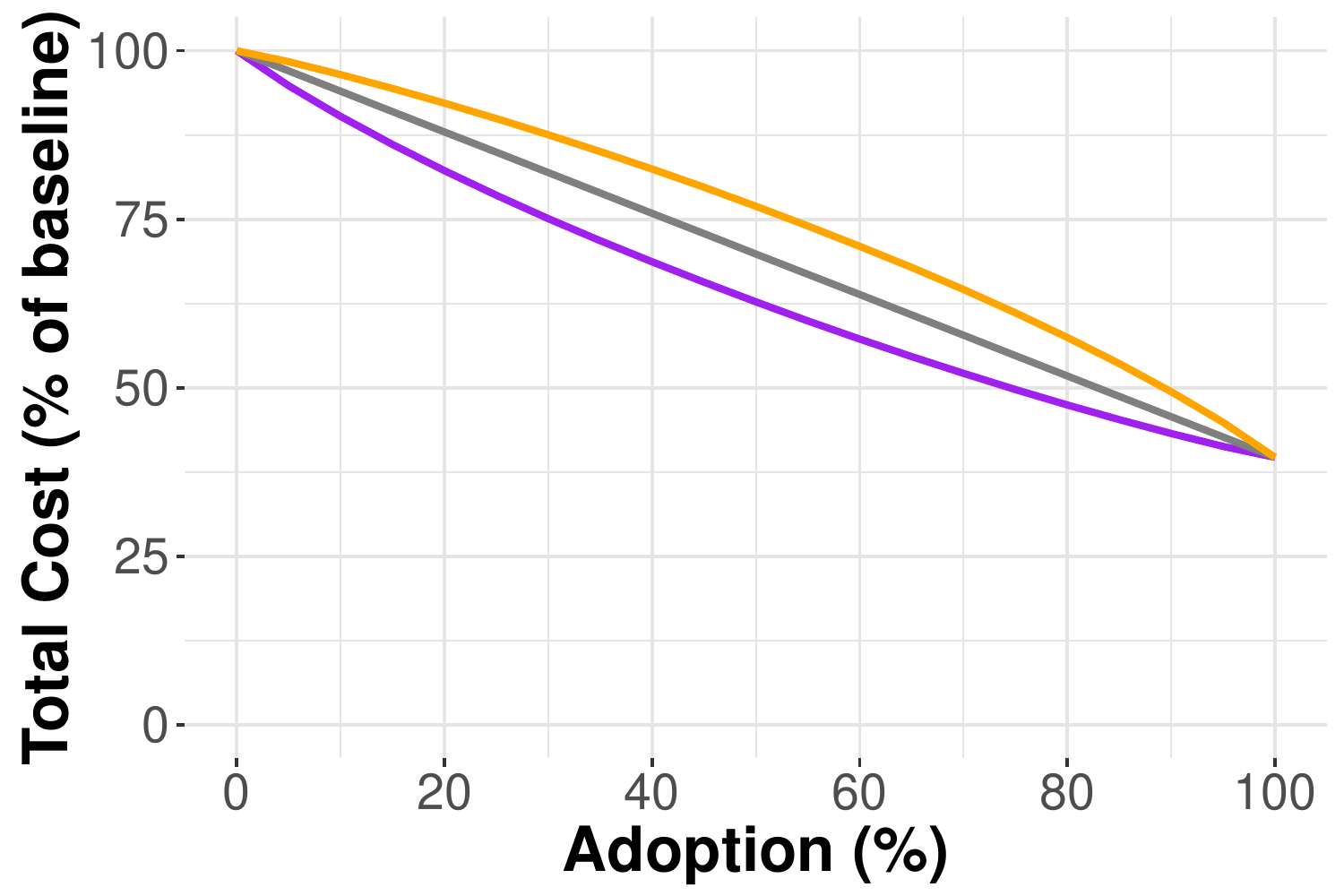}
\caption{The total annual cost of electricity without coordination for all households, as a percent of the baseline total cost without technology ($T_{BL}$), is plotted against the percent of households that have adopted storage and PV, using normalized savings as the adoption criterion. This analysis is under Policy 1 prices. The purple curve is for the forward adoption pattern, the orange curve is for reverse adoption, and the gray curve is for random adoption.
The reduction in total annual cost is the sum of the absolute annual savings of adopters, and absolute savings are modestly positively correlated with normalized savings. Thus, under the forward adoption pattern, the first adopters tend to have higher absolute savings than those under the reverse adoption pattern. That is why the forward adoption curve lies below the reverse adoption curve, with the random adoption curve in between. $T_{BL}$ is about \$903 million, and it serves as the baseline cost used for Figs. \ref{fig:Value_of_Coord_no_noise} through \ref{fig:Value_of_Coordination_Multi}. Note that even at 100\% adoption of net-zero systems, the households together reduce their costs only by 60\%. This is due to the combination of the gap between the purchase price and the sale price and the mismatch between the timing of their consumption and their generation, which their storage devices can only partially mitigate.}
\label{fig:Total_Cost_Curves}
\end{figure}

Define the total cost under forward adoption as $T_{\textrm{fwd}}(t) = \sum_{i\in A_{\textrm{fwd}}(t)}b_{N,i} + \sum_{i\in B_{\textrm{fwd}}(t)}b_{BL,i}$.
In other words, add up the new bills (i.e., with technology) for the adopters and the baseline bills (i.e., without technology) for the non-adopters.
The total costs under reverse adoption and random adoption are defined analogously.
Next, define the baseline total cost without technology as $T_{BL}=\sum_{i=1}^{N}b_{BL,i}$.
Figure \ref{fig:Total_Cost_Curves} plots $T_{\textrm{fwd}}(t)$, $T_{\textrm{rev}}(t)$, and $T_{\textrm{rdm}}(t)$ as a percent of $T_{BL}$, under Policy 1 pricing.

\subsubsection{Value of Coordinated Action}

The group minimizes its electricity costs on a daily basis.
For a given day, we set up the optimization as follows.
Note that the VCA is under perfect foresight.

We take the mean of prices across all households to be the prices the coordinator faces.
That is, $\textbf{q}_G^{(j)}=\frac{1}{N}\sum_{i=1}^N \textbf{q}_i^{(j)}$, and $\textbf{r}_G^{(j)}=\frac{1}{N}\sum_{i=1}^N \textbf{r}_i^{(j)}$.

Take a given level of adoption, $t$\%, and assume the forward adoption pattern.
The solar energy generated by the group, $\textbf{e}_G^{(j)}$, is that generated by the PV systems of the adopters: $\textbf{e}_G^{(j)} = \sum_{i\in A_{\textrm{fwd}}(t)}\textbf{e}_i^{(j)}$. The group's inflexible load is the total across all households: $\textbf{l}_G^{(j)} = \sum_{i=1}^N \textbf{l}_i^{(j)}$.
The group's net load $\textbf{n}_G = \textbf{l}_G^{(j)}-\eta_I\textbf{e}_G^{(j)}$.

We obtain the group's storage device capacity and charging and discharging rates by summing across households: $\underline{u}_G=\sum_{i\in A_{\textrm{fwd}}(t)}\underline{u}_i$, $\overline{u}_G=\sum_{i\in A_{\textrm{fwd}}(t)}\overline{u}_i$, and $\overline{x}_G=\sum_{i\in A_{\textrm{fwd}}(t)}\overline{x}_i$.

The coordinator solves the following optimization problem to schedule the operation of the collective storage.

\begin{subequations}
\label{eqn:Optimization_Group}
\begin{align}
\underset{\textbf{u}_G^{(j)},\textbf{x},\textbf{g}_G \; \in \; \mathbb{R}^{24}}{\text{minimize}} \; & [\textbf{g}_G]_+^T\textbf{p}_b + [\textbf{g}_G]_-^T\textbf{p}_s \\
\mbox{subject to} \; & \textbf{g}_G=\textbf{n}_G+\frac{1}{\eta_C\eta_I}[\textbf{u}_G^{(j)}]_++(\eta_D\eta_I)[\textbf{u}_G^{(j)}]_- \\
& -\underline{u}_G \leq \textbf{u}_G^{(j)} \leq \overline{u}_G \\
& 0 \leq \textbf{x} \leq \overline{x}_G \\
& \textbf{x}_0 = x_0 \\
& \textbf{x}_{h} = \textbf{x}_{h-1}+\textbf{u}_{G,h}^{(j)},\; \forall h \in \lbrace1,\ldots,24\rbrace,
\end{align}
\end{subequations}

The coordinator then controls the adopters' storage devices to execute the optimal schedule $\textbf{u}^{(j)\star}_G$ at the aggregate level.
There are generally many ways to distribute $\textbf{u}^{(j)\star}_G$ over the adopters' devices.

An implicit assumption of this model of coordination is that the group of households has free use of the wires between the homes and that there are no losses in sharing energy this way.
Surplus energy supplied by one home in the group, whether by its rooftop PV or its storage device, can be used to offset the consumption of a separate home in the group with no loss and no fees, other than the losses induced by the first home's inverter and storage device inefficiencies.
In effect, this model of coordination treats the group of homes as one big household, combining their consumption, rooftop PV generation, and storage devices.

The cost of electricity for the group of households for this day will be the optimal value of the objective; call it $c_G^{(j)}$.
Define $C_{fwd}(t)=\sum_{j=1}^{Y}c_G^{(j)}$.
The VCA for the forward adoption pattern at adoption level $t$\% is then $\textrm{VCA}_{\textrm{fwd}}(t)=T_{\textrm{fwd}}(t)-C_{\textrm{fwd}}(t)$.
The VCA for the reverse and random adoption patterns are defined analogously.
Figure \ref{fig:Value_of_Coord_no_noise} plots the VCA for different adoption patterns and levels.

\subsubsection{Value of Coordinated Information}

The VCI is defined as the additional savings achieved by coordination on top of the VCA in the presence of forecast errors.

For a given day, the coordinator must generate forecasts $\hat{\textbf{l}}^{(j)}_G$ and $\hat{\textbf{e}}^{(j)}_G$.
Let $\hat{\textbf{l}}_{G,h}^{(j)} = [ \textbf{l}_{G,h}^{(j)}+\epsilon_{G,h}^{(j)}]_+$, and $\hat{\textbf{e}}_{G,h}^{(j)} = [ \textbf{e}_{G,h}^{(j)}+\gamma_{G,h}^{(j)}]_+$ where $\epsilon_{G,h}^{(j)}$ and $\gamma_{G,h}^{(j)}$ are independently and identically distributed across $h$ and $j$.
The coordinator is forecasting aggregate quantities, so its error is lower than that of the individual household.
$\epsilon_{G,h}^{(j)}$ is normally distributed with $\mu=0$ and $\sigma=w(\bar{\textbf{L}}_G)\bar{\textbf{L}}_G$, where $\bar{\textbf{L}}_G=\frac{1}{N}\sum_{i=1}^N\bar{\textbf{L}}_i$, and $w(\cdot)$ is the fit for the CV scaling law for forecasting model $\mathcal{M}_3$ from \cite{sevlian2018}. $\gamma_{G,h}^{(j)}$ is normally distributed with $\mu=0$ and $\sigma=w(\bar{\textbf{E}}_G)\bar{\textbf{E}}_G$, where $\bar{\textbf{E}}_G=\frac{1}{N}\sum_{i=1}^N\bar{\textbf{E}}_i$.

Let $\hat{\textbf{n}}_G = \hat{\textbf{l}}^{(j)}_G-\eta_I\hat{\textbf{e}}^{(j)}_G$.
The coordinator solves the same linear program as before, except that $\hat{\textbf{n}}_G$ replaces $\textbf{n}_G$ in constraint (\ref{eqn:Optimization_Group}b).
The optimal schedule $\textbf{u}_G^{(j)\dagger}$ from this solution is the coordinator's schedule of charging and discharging for the collective storage.

Let $\textbf{g}_G^{\dagger}=\textbf{n}_G+\frac{1}{\eta_C\eta_I}[\textbf{u}_G^{\dagger}]_++(\eta_D\eta_I)[\textbf{u}_G^{\dagger}]_-$, where $\textbf{n}_G$ is the true net load, i.e. $\textbf{n}_G=\textbf{l}^{(j)}_G-\eta_I\textbf{e}^{(j)}_G$.
The group's cost of electricity for this day is $c_G^{\dagger(j)}=[\textbf{g}_G^{\dagger}]_+^T\textbf{q}_G^{(j)} + [\textbf{g}_G^{\dagger}]_-^T\textbf{r}_G^{(j)}$.
Let $C_{\textrm{fwd}}^{\dagger}(t)=\sum_{j=1}^{Y}c_G^{\dagger(j)}$.
The VCI for the forecast error CV of $P$ and adoption level of $t$\% is then:
\begin{equation}
\textrm{VCI}_{\textrm{fwd}}(t)=\sum_{i\in A_{\textrm{fwd}}(t)}b^{\dagger}_{N,i}(P) + \sum_{i\in B_{\textrm{fwd}}(t)}b_{BL,i}-C_{\textrm{fwd}}^{\dagger}(t)-\textrm{VCA}_{\textrm{fwd}}(t).
\end{equation}
The VCI for the reverse and random adoption patterns are defined analogously.
Figure \ref{fig:VCI} plots the VCI for different adoption patterns, forecast error CV levels, and adoption levels.

\subsection{Data availability}

The pricing and solar irradiance data are publicly available.
The smart meter data is the property of PG\&E and was obtained by the authors under a non-disclosure agreement.

\bibliographystyle{IEEEtran}
\bibliography{bibtex/bib/Coord_paper}

\end{document}